\documentclass[conference]{IEEEtran}
\IEEEoverridecommandlockouts
\usepackage{cite}
\usepackage{url}
\usepackage{amsmath,amssymb,amsfonts}
\usepackage{algorithmic}
\usepackage{graphicx}
\usepackage{textcomp}
\usepackage{xcolor}
\usepackage{subfigure}
\usepackage[linesnumbered,algoruled,boxed,lined]{algorithm2e}
\usepackage{fancyhdr}

\SetKwInOut{Parameter}{parameter}
\def\BibTeX{{\rm B\kern-.05em{\sc i\kern-.025em b}\kern-.08em
    T\kern-.1667em\lower.7ex\hbox{E}\kern-.125emX}}
\begin{document}

\pagestyle{fancy}
\fancyhf{} 
\fancyhead[C]{\small This article has been accepted for publication in IEEE Transactions on Network and Service Management. This is the author's version which has not been fully edited and
content may change prior to final publication. Citation information: DOI 10.1109/TNSM.2024.3435505.} 
\fancyfoot[C]{\thepage} 

\title{PHaul: A PPO-based forwarding agent for Sub6 enhanced Integrated Acess and Backhaul networks\\
\thanks{This work was funded by the European Commission through the SNS JU project NANCY (grant agreement No. 101096456). The authors acknowledge CERCA Programme/Generalitat de Catalunya for sponsoring part of this work.}
}

\author{\IEEEauthorblockN{Jorge Pueyo}
\IEEEauthorblockA{\textit{I2CAT Foundation} \\
jorge.pueyo@i2cat.net}
\and
\IEEEauthorblockN{Daniel Camps-Mur}
\IEEEauthorblockA{\textit{I2CAT Foundation} \\
daniel.camps@i2cat.net}
\and
\IEEEauthorblockN{Miguel Catalan-Cid}
\IEEEauthorblockA{\textit{I2CAT Foundation} \\
miguel.catalan@i2cat.net}
}

\maketitle
\thispagestyle{fancy} 

\begin{abstract}
3GPP Integrated Access and Backhaul (IAB) allows operators to deploy outdoor mm-wave access networks in a cost-efficient manner, by reusing the same spectrum in access and backhaul. In IAB networks the performance bottleneck is the wireless backhaul segment, where efficient forwarding strategies are needed to effectively use the available capacity. In addition, the performance of the mm-wave IAB backhaul segment is contingent on the availability of line of sight (LoS) conditions in the selected deployment sites. To mitigate LoS dependence, in this paper, we propose to complement the mm-wave backhaul segment of IAB networks with additional Sub6 backhaul links, which contribute to the capacity and robustness of the backhaul network. We refer to IAB networks combining Sub6 and mm-wave links in the backhaul as Sub6 enhanced IAB networks. In this context, the main contribution of this paper is PHaul, a forwarding engine for Sub6 enhanced IAB networks that accomodates different traffic engineering criteria, and combines an offline path selection heuristic with an online Deep Reinforcement Learning (DRL) agent based on Proximal Policy Optimization (PPO). By leveraging a network digital twin of the IAB wireless backhaul, PHaul periodically samples the input traffic of the backhaul network and updates flow to path mappings, with execution times below 10 seconds in realistic backhaul topologies. We present an exhaustive performance evaluation, where we demonstrate that PHaul can achieve gains of up to 36\% in throughput efficiency and of up to 20\% in fairness, when compared against two alternative heuristics in a wide range of network configurations. We also demonstrate that PHaul is robust to differences between the network topologies considered in the training and inference phases, which can occur in practice due to link failures.

\end{abstract}


\section{Introduction}
\label{sec:intro}
Addressing the data rate demands of future communication services will require heterogeneous 5G-Advanced (5G-A) networks with a dense high-capacity layer composed of outdoor mm-wave (26/28 GHz) small cells \cite{6g_reqs}. This trend is expected to continue in 6G networks, where the Sub-THz band (90GHz -300GHz) is being considered to deliver even greater access capacities \cite{6g_thz}.

Laying out dedicated fibre for backhauling is a roadblock to enabling a massive deployment of outdoor small cells. To address this challenge 3GPP introduced the Integrated Access and Backhaul (IAB) technology in Release 16 \cite{TS38300}. The idea of IAB is to enable an operator to start by deploying a small number of fibre connected mm-wave small cells, known as \emph{donor} nodes in IAB terminology, and to extend service coverage by deploying additional \emph{IAB nodes} that are wirelessly backhauled through donor nodes, and can provide direct service to User Equipments (UEs), or act as parent nodes for other IAB nodes. When service demand grows, an operator can add fibre connectivity to IAB nodes, upgrading them to donor nodes, thus increasing the effective backhaul capacity in a cost-efficient manner.

While the IAB model defined by 3GPP is spectrum agnostic and can operate either in the above-6 GHz or sub-6 GHz spectrum, \textcolor{black}{the main use case for IAB networks is to provide mm-wave spectrum in the access and backhaul, complementing the Sub6 spectrum offered by the macro-cell layer.} However, mm-wave backhauling in urban environments is often subject to link blocking issues, and to non-line of sight performance degradation \cite{mmwave-issues}. Therefore, in this work we consider combining Sub6 spectrum together with mm-wave frequencies in the IAB backhaul segment to address both coverage and capacity requirements in dense urban deployments of small cells. \textcolor{black}{Limiting the use of Sub6 in IAB to the backhaul segment reduces potential cross-interference with the macro-layer, because backhaul links are fixed and can be engineered to avoid interference. In addition, the upper part of the 6 GHz band, which was recently authorized for IMT use in the WRC23 conference \cite{wrc23}, could be exploited for the IAB backhaul, avoiding any interference with the macro layer.} We refer to IAB networks combining Sub6 and mm-wave links in the backhaul as Sub6 enhanced IAB networks.

When planning a traditional wireless backhaul network, a common approach is to overprovision based on busy hour. However, this is not a good approach for IAB networks in urban environments, for the following reasons. First, the limited coverage of the mm-wave access results in quick variations in the number of UEs connected to a given cell, which complicates the estimation of the required backhaul capacity. Second, overprovisioning the IAB backhaul is not economically efficient as it removes away access capacity, thus requiring the deployment of additional IAB nodes. Hence, if overprovisioning is not a practical approach for planning IAB networks, it is critical to consider flexible routing schemes that can maximize the efficiency of the wireless backhaul by adapting backhaul paths to the available load demands at a given time.

To route packets in IAB-based networks, 3GPP has defined the \emph{Backhaul Adaptation Protocol} (BAP) that is based on source routing. Namely, a set of paths between each IAB node and its target donor node are pre-computed and the next hop entry for each path is programmed into the forwarding engine of each IAB node. Thus, higher layer packets arriving at an IAB node are matched to their corresponding path according to their intended destination. This source-based routing model can be easily extended to a heterogeneous backhaul network that comprises both mm-wave and Sub6 frequencies, including within the set of available paths between an IAB node and a donor, both Sub6 and mm-wave paths. BAP can be used both in tree and mesh topologies, where mesh topologies can be supported in IAB by having an IAB node connect to more than one IAB parent node, for example by including multiple IAB Mobile Termination (MT) functions in the same node, or by means of using Dual Connectivity (DC) \cite{TS38300}.

Under the BAP forwarding model, even though paths are precomputed, the classifier matching higher layer flows into the pre-provisioned backhaul paths can be dynamically updated.
The main focus of this paper is to investigate mechanisms that dynamically update the matching between backhaul flows and available paths in Sub6 enhanced IAB backhaul networks, to optimize a given traffic engineering criteria. In particular, we investigate the use of Deep Reinforcement Learning (DRL) to design a forwarding agent that, once trained on a given IAB backhaul topology, is able to dynamically update the flow to path mappings according to varying traffic demands.

The main contribution of this paper is threefold. 
\begin{itemize}
\item [i.] First, we present a novel IAB architecture that combines Sub6 and mm-wave links in the backhaul segment.
\item [ii.] Second, we present PHaul, an IAB forwarding engine that combines an offline path selection heuristic with an online DRL agent based on Proximal Policy Optimization (PPO) \cite{ppo}. PHaul supports flexible traffic engineering criteria and leverages a network digital twin to perform path allocations in Sub6 enhanced IAB networks.
\item [iii.] We describe the training process of PHaul, and we perform an exhaustive simulation-based evaluation comparing PHaul to two alternative greedy algorithms, both in terms of performance and inference time.
\item [iv.] \textcolor{black}{Finally, to ease reproducibility, we have open sourced our implementation of PHaul\footnote{\textcolor{black}{https://github.com/Fundacio-i2CAT/phaul/}}, as well as the environments we have created to evaluate its performance.}
\end{itemize}

The paper is structured as follows. Section \ref{sec:related_work} describes related work. Section \ref{sec:network_model} presents the considered network model, Section \ref{sec:agent_design} describes the design of PHaul, and Section \ref{sec:perf_eval} evaluates the performance of PHaul against potential alternatives. Finally, Section \ref{sec:conclusions} summarizes and concludes the paper.

\section{Related Work}
\label{sec:related_work}
We survey the state of the art relevant to this work grouped in: i) works that explore the utilization and combination of Sub6 and mm-wave bands for wireless backhauling, ii) works that explore the use of Machine Learning (ML) mechanisms for routing, and iii) works that focus on flow routing in IAB networks.

Although usually focused on mm-wave bands \textcolor{black}{and fixed IAB nodes}, several works \textcolor{black}{propose novel architectures to enhance IAB performance in specific scenarios}. Authors in \cite{sub6-ref1} present and evaluate through simulations an IAB architecture using Sub6 bands to support wide coverage and mobile systems. In \cite{sub6-ref2} the authors consider Sub6 IAB to introduce a novel solution to allocate access and backhaul resources. The solution combines centralized (IAB donor) and distributed (IAB nodes) approaches to handle both non-bursty and bursty traffic, respectively. \textcolor{black}{Authors in \cite{ris-ref} introduce a dynamic resource management framework for UAV-assisted IAB networks utilizing Reconfigurable Intelligent Surfaces (RIS). This framework uses a distributed Stackelberg game-theoretic approach to optimize end-to-end energy efficiency by exploiting the integration of UAVs with IAB and RIS technologies.}

Regarding the combination of Sub6 and mm-wave bands in the wireless backhaul, the authors in \cite{5gxhaul-bristol} demonstrate a city-wide deployment that combines IEEE 802.11ad radios in the 60GHz band, with IEEE 802.11 radios in the 5 GHz band. This approach however is not based on IAB. Instead, an SDN-based control plane is used to implement source-based routing. In \cite{sodalite_tnsm} the authors present a centralized SDN-based wireless backhaul system, also based on IEEE 802.11 radios in the 5 GHz and 60 GHz bands, which allocates for each flow a main and a backup path using heuristic policies that search across a set of $K$ predefined paths between each source-destination pair. The authors do not report on the execution times of their proposed heuristic. 

\begin{table*}[!t]
\centering
\caption{\textcolor{black}{Summary of the state of the art}}
\label{tab:soa}
\begin{tabular}{|p{1.5cm}|p{3.5cm}|p{6cm}|p{5cm}|}
\hline
\textbf{Reference Number} & \textbf{Main contribution} & \textbf{Routing/Scheduling solution}  & \textbf{Performance evaluation} \\
\hline
\centering
\cite{sub6-ref2} & IAB network (only sub6), avoids conflict in resource allocation for bursty-traffic & Centralized donor allocates hard resources (non-bursty traffic), IAB nodes apply distributed allocation (bursty- traffic)  & Simulation: GBR-only and bursty traffic scenarios\\
\hline
\centering
\cite{ris-ref} & Integrates RIS and UAV-assisted IAB, targets energy efficiency & Leader UAV determines RIS configuration, bandwidth splitting, UAV Tx power and initial UE Tx Power. UEs determine optimal power according to bandwidth. Stackelberg game-based iteration. & Modeling: Convergence of the optimisation process. Simulation: single UAV, RIS and BS, comparison of energy-efficiency vs data rate approaches \\
\hline
\centering
\cite{sodalite_tnsm} & SDN-based Wi-Fi backhaul, centralized per-flow path computation plus distributed fast-recovery when links fail & Computes a subset of paths per flow, selects main and backup path according to network status and flow requirements, creates forwarding rules to apply fast local link reroute & Modeling: analysis of number of rules and signaling overhead. Simulation: performance evaluation according to network size. Experimental: Link Failure detection time, E2E evaluation in a real testbed\\
\hline
\centering
\cite{r2l} & Wired network, per packet decision on next hop, centralized RL agent & RL based on neural network and evolutionary algorithm. State: local network measurements (e.g. queue size) and packets header. Action: outgoing port. Reward: configurable (e.g. delay or throughput) & Simulation: fixed number of nodes (5 or 16), comparison with shortest path, longest path and load balancing approaches\\ 
\hline
\centering
\cite{qrouting} & Wired network, per packet decision on next hop, one RL agent per router & RL based on Q-learning. State: N-Optimal paths, topology, link state, delay measurements. Action: path selection. Reward: latency  & Simulation: Comparison with other approaches in terms of scalability, robustness and delay.  \\
\hline
\centering
\cite{recce} & Industrial IoT networks, joins routing and TDMA scheduling, centralized RL agent & RL based on PPO. State: State of transmissions in each node at a given slot (queue length, time remaining, hops remaining, priority), precomputed paths. Action: Chooses transmissions in the next slot (according to 6 criteria) and their next hop. Reward: delay value weighted by deadline delay and delivery status & Simulations: impact of number of routes, channels and nodes. Performance in terms of packet delivery ratio, impact of untrained topologies\\
\hline
\centering
\cite{iab_routing} & IAB network (only mmWave), joins routing and resource allocation, one RL agent per IAB node &  RL based on Soft Actor-Critic algorithm. State: information from neighbor nodes (channel condition, latency, reliability). Action: next hop in DL and UL. Reward: Combination of latency and reliability. &  Simulations: 1 donor, 18 IAB nodes, variable number of UEs. Computational complexity. Performance in terms of average delay and reliability \\
\hline
\centering
Phaul & IAB network (mmwave and sub6), per-flow path allocation, centralized RL agent & RL based on PPO. State: Precomputed N-optimal paths (least common heuristic) and input data rate per flow, current path and measured effective data rate (digital twin). Action: Path selection. Rewards: Configurable according to throughput efficiency and fairness &  Simulations: 3 donors, up to 60 IAB nodes, variable topologies and offered load. Impact of path computation heuristics, training steps, broken links and untrained topologies. Performance in terms of throughput efficiency and fairness.\\
\hline
\end{tabular}
\end{table*}

The application of ML-based techniques for routing is motivated by the large computational costs of applying optimal routing strategies. An optimal solution to the path allocation problem with traffic engineering constraints, given a known topology and demand matrix, is described in \cite{MCFP} through a Multi-Commodity Flow Problem (MCFP) formulation where flows in each link share bandwidth using a max-min fairness criteria. This method is shown to take as long as one hour for some problem instances, thus not being practical in wireless backhaul networks with constantly changing demands. To achieve lower execution times, recent works in the state of the art have started to study the application of Machine Learning (ML) techniques. Most of these works adopt a Reinforcement Learning (RL) framework, where a routing agent is considered that selects a path or next-hop as action, models the environment using counters available in routers or switches, and defines a reward that captures a particular routing objective. 

In \cite{r2l}, R2L, a centralized routing agent, is presented that takes a next-hop decision on each new packet arrival at a wired router. The agent can be trained to optimise different criteria (e.g. throughput, delay), and network state is built from the queue lengths at different routers. The agent is shown to converge in the training phase, although only a simplified network topology is considered in the evaluation. In \cite{qrouting} a traditional Q-learning algorithm is used to select among a set of $N$ paths between each source and destination in a wide area network. The agent is trained with the objective of choosing the path with lowest delay. Closest to our work is the RECCE agent proposed in \cite{recce}, which however focuses on a different use case, namely industrial WirelessHart IoT networks. RECCE is based on a modern Proximal Policy Optimization (PPO) agent and performs joint routing and scheduling by choosing among a set of $\gamma$ precomputed paths and among six different scheduling strategies. A single reward is proposed to maximize the number of packets delivered within their deadline. 

Looking at routing for IAB networks the authors in \cite{iab_routing} propose a DRL scheme for both routing and resource allocation. In their proposal each IAB node embeds an independent DRL agent that selects the next next hop in uplink and downlink. Unlike this proposal, in this paper we propose a global DRL agent that performs routing decisions considering the network as a whole. In \cite{iab_planning} the authors propose a genetic algorithm to assit the planning stage of IAB networks, and a routing heuristic to cope with temporary link blockages. This paper though does not address the problem of load balancing in IAB networks, which is the main focus of our work.

To the best of our knowledge, our work is the first to address the problem of ML-based routing in Sub6 enhanced IAB networks. \textcolor{black}{Table \ref{tab:soa} summarizes the main features of the solutions presented in this section, providing a comparison with our Phaul proposal.}

\section{Network Model}
\label{sec:network_model}

\begin{figure*}[!t]
  \centering
  \includegraphics[width=0.9\linewidth]{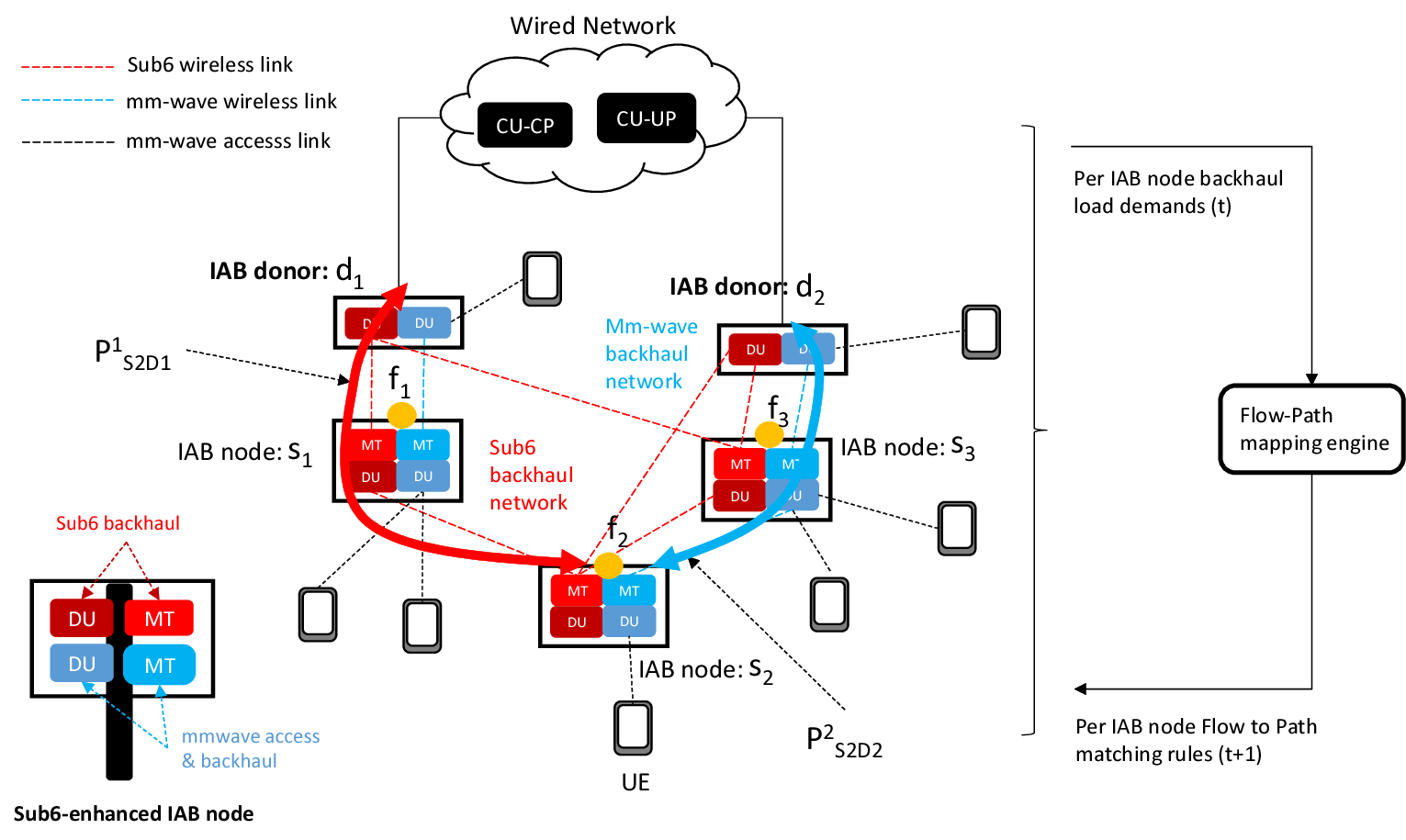}
  \caption{PHaul network model}
  \label{fig:network_model}
\end{figure*}

Figure \ref{fig:network_model} depicts the network model considered in this work, consisting of an IAB network where the backhaul segment is enhanced with additional Sub6 backhaul links. In the bottom left of Figure \ref{fig:network_model} we depict the logical architecture of a Sub6-enhanced IAB node, which, following the IAB architecture, features a Distributed Unit (DU) and Mobile Terminated (MT) function both for the Sub6 and the mm-wave bands. However, only the mm-wave DU will be used to connect UEs in the access, as Sub6 access coverage is already provided by the macro layer. 

The central part of Figure \ref{fig:network_model} depicts the following elements:  

\begin{itemize}
    \item[i.] IAB donor nodes, denoted as $d_{k} \in \mathbb{D}$, featuring both a Sub6 and a mm-wave DUs, which connect UEs and other IAB nodes to the wired network, where the Centralized Unit (CU) components are located.
    
    \item[ii.] IAB nodes, denoted as $s_{i} \in \mathbb{S}$, which provide access to UEs and other IAB nodes. An IAB node may directly connect to an IAB donor or to another IAB node. Multi-path is supported by embedding multiple MT functions into a single IAB node or by using Dual Connectivity, and allows an IAB node to establish more than one link to other IAB donor or IAB parent nodes. IAB nodes provide access to multiple UEs, where each UE establishes a separate PDU session. However, we consider that in the backhaul, traffic from all UEs is aggregated into a single backhaul PDU session, which we refer to as a \emph{backhaul flow} $f_{i} \in \mathbb{F}$.
    
    \item[iii.] Backhaul flows, $f_{i} \in \mathbb{F}$, originate at non-donor IAB nodes, i.e. $s_{i} \in \mathbb{S}$. 
    Each backhaul flow carries a time varying traffic load, $\lambda_{i}(t)$, generated by all the UEs in RRC\_CONNECTED state. Each backhaul flow is mapped to a pre-provisioned IAB \emph{backhaul path} defined between an IAB node $s_{i} \in \mathbb{S}$ and an IAB donor $d_{i} \in \mathbb{D}$. For each flow $f_{i}$, the source is fixed, i.e. $s_{i}$, however any donor $d_{k}$ could be used as destination, as long as a backhaul path exists. The reason is that all donor nodes provide access to the wired network where the CU resides. 
    
    \item[iv.] The Sub6 backhaul, shown in red in Figure \ref{fig:network_model}, and the mm-wave backhaul, shown in blue in Figure \ref{fig:network_model}, provide a set of backhaul paths between IAB nodes and IAB donors. Due to different propagation characteristics in the Sub6 and mm-wave bands, different paths may be available in the Sub6 and mm-wave backhaul networks. The BAP protocol in IAB uses source-based forwarding. Hence, each flow $f_{i} \in \mathbb{F}$ is bound to use a single path, where paths are pre-provisioned. In particular, we consider that a set of $k \leq K^{max}$ paths are pre-provisioned for each flow in the Sub6 and the mm-wave backhaul networks, where $P^{n}_{s_{i},d_{k}}$ indicates the $n$-th pre-provisioned path for flow $f_{i}$ between $s_{i}$ and $d_{k}$. Thus, a given backhaul flow could be routed across a total of $2K^{max}$ paths considering both networks, but can only use a single path at a given time.     
    
    \item[v. ] Finally, a flow-path mapping engine is defined as a control plane entity that periodically obtains the state of the network, e.g., through reading load counters in the IAB nodes. Based on the obtained information, the flow-path mapping engine updates the mappings between backhaul flows and backhaul paths in each IAB node. The PHaul agent resides in the flow-path mapping engine.
\end{itemize}

To model the capacity available in the wireless backhaul, we consider a fixed link capacity and neglect cross-link interference. This assumption is possible in the case of IAB mm-wave links by allocating dedicated slots to the backhaul links and employing beamforming to avoid cross-link interference \cite{TS38300}. In the case of the Sub6 backhaul links, we consider the use of the 6GHz band, where 29 channels of 40 MHz are available \cite{fcc_6ghz}, without legacy Wi-Fi devices, which can be used to avoid cross-link interference in practical urban wireless backhaul networks. The exact capacities of the Sub6 and mm-wave links depend on the specific radio configuration, 
which will be described in detail in Section \ref{sec:perf_eval}.

Notice that various backhaul flows can traverse the same backhaul node and compete for the capacity of a given Sub6 or mm-wave backhaul link. In this case, we assume that backhaul nodes assign capacity to each flow traversing a backhaul link using a max-min fairness criteria \cite{maxmin}, which is a good approximation when competing flows through a bottleneck carry TCP traffic \cite{tcp_mmf}.    

Based on the described network model we can define the set of paths available to flow $f_{i} \in \mathbb{F}$ as $\mathbb{P}^{selected}_{i}$, with $|\mathbb{P}^{selected}_{i}| \leq 2K^{max}$. Periodically, the flow-path mapping engine samples the traffic matrix from each backhaul flow and updates the per-flow path allocations. The goal of PHaul is to, based on the current traffic matrix $\{ \lambda_{i}(t) \}$, assign for each flow $f_{i} \in \mathbb{F}$ a single path $P^{opt}_{s_{i},d_{i}} \in \mathbb{P}^{selected}_{i}$, to optimize a given traffic engineering criteria $J^{TE}$. The execution time of the PHaul agent is the key factor to determine how often path allocations can be adapted to the varying traffic matrices.

Regarding the traffic engineering criteria, we consider $J^{TE}$ to be a varying objective function defined by the operator of the IAB network as a function of the effective data rates, $\phi_{i}(t)$, allocated to each backhaul flow, where the effective data rate represents the actual rate that can be served from that flow, as compared to the overall flow demand represented by $\lambda_{i}(t)$. 

A key requirement in the design of PHaul is to be able to operate with any traffic engineering criteria defined by the network operator. For example, in this work we consider the following $J^{TE}$ criteria:

\begin{itemize}
\item[i.] \emph{Throughput efficiency}:
\begin{equation}
    0 \leq J^{TE} = \frac{\sum_{i} \phi_{i}(t)}{\sum_{i} \lambda_{i}(t)} \leq 1
\end{equation}
where the goal is to maximize the proportion of load that can be carried by the network. \textcolor{black}{This traffic engineering criteria would be appropriate in networks where demand is expected to be heterogeneous across IAB nodes. For example, a city center where some IAB nodes are covering a tourist hotspot, whereas other nodes cover more quiet areas.}
\item[ii.] \emph{Fairness}:
\begin{equation}
    0 \leq J^{TE} = \frac{(\sum_{i} \phi_{i}(t))^{2}}{|\mathbb{F}|\sum_{i} \phi_{i}(t)^{2}} \leq 1
\end{equation}
where $|\mathbb{F}|$ is the number of flows, and the goal is to maximize fairness across flows. \textcolor{black}{This traffic engineering criteria would be appropriate when demand across IAB nodes is expected to be uniform, for example when an IAB network is used to cover a suburban area.}
\item[iii.] \emph{Weighted}:
\begin{equation}
    0 \leq J^{TE} = (1 + \gamma) \frac{\sum_{i} \phi_{i}(t)}{\sum_{i} \lambda_{i}(t)} + (1 - \gamma)\frac{(\sum_{i} \phi_{i}(t))^{2}}{|\mathbb{F}|\sum_{i} \phi_{i}(t)^{2}} \leq 2
\end{equation}
\textcolor{black}{where the the parameter $-1 < \gamma \leq 1$ allows the operator to balance throughput and fairness according to its business goals. Being the more general option, we use this traffic engineering criteria with different values of $\gamma$ in the rest of the paper.}
\end{itemize}

\section{PHaul design}
\label{sec:agent_design}
\subsection{Design Principles}
As introduced in Section \ref{sec:network_model}, the goal of PHaul is to perform per-flow path allocations that optimize a traffic engineering criteria, $J^{TE}$, with execution times in the order of few seconds to track variations in the IAB access traffic. As described in Section \ref{sec:related_work}, traditional optimization methods cannot be applied to this problem as they lead to excessive computational times. To address this problem, we turn our attention to Deep Reinforcement Learning (DRL). 

The goal of the PHaul DRL agent is to periodically collect traffic demands for all flows, i.e. $\{ \lambda_{i}(t) \}$, and make a path allocation decision, i.e. select $P^{opt}_{s_{i},d_{i}} \in \mathbb{P}^{selected}_{i}$, $\forall f_{i} \in |\mathbb{F}|$. Given the network model introduced in Section \ref{sec:network_model}, a naive DRL implementation that jointly allocates all flows in each action, results in an action space of size $|\mathbb{A}|=(2K^{max})^{|\mathbb{F}|}$, where $|\mathbb{F}|$ is the number of flows, and $K^{max}$ the number of available paths per flow in the Sub6 and mm-wave networks. This approach is not feasible for practical networks, as the action space can quickly become too large. 

The main insight of PHaul is an approach to reduce the size of the action space, while still being able to allocate a path for each flow. The idea is to define a \emph{digital twin} of the IAB backhaul network over which we can define a reduced action space, which consists of the allocation of a path for a single flow, instead of a joint allocation for the $|\mathbb{F}|$ flows. The agent then observes the resulting reward of the applied action over the digital twin, and continues sampling the reduced action space until an allocation for all the flows is obtained, which can then be programmed over the real network. The rationale behind this design approach is the following:
\begin{itemize}
\item[i.] The topology and link characteristics of the Sub6 enhanced IAB network are expected to be stable, and can therefore be easily reproduced by a digital twin (e.g. a network simulator). If the topology changes significantly, e.g. due to link failure, then the digital twin can be correspondingly updated.
\item [ii.] The traffic demands $\{ \lambda_{i}(t) \}$, can also be considered stable for periods of several seconds and can therefore be modelled in the digital twin. This is the time budget available to perform an allocation decision.
\item[iii.] Reducing the action space, i.e. allocating a path for only one flow at each action, simplifies training and leads to better convergence properties of the DRL agent.
\item[iv.] The repeated application of the simplified agent over the digital twin should allow to sample the larger action space leading to well performing allocations.
\end{itemize}
In Section \ref{sec:perf_eval} we validate empirically whether the previous assumptions lead to a well performing forwarding agent.

\begin{figure}[ht]
  \centering
  \includegraphics[width=\linewidth]{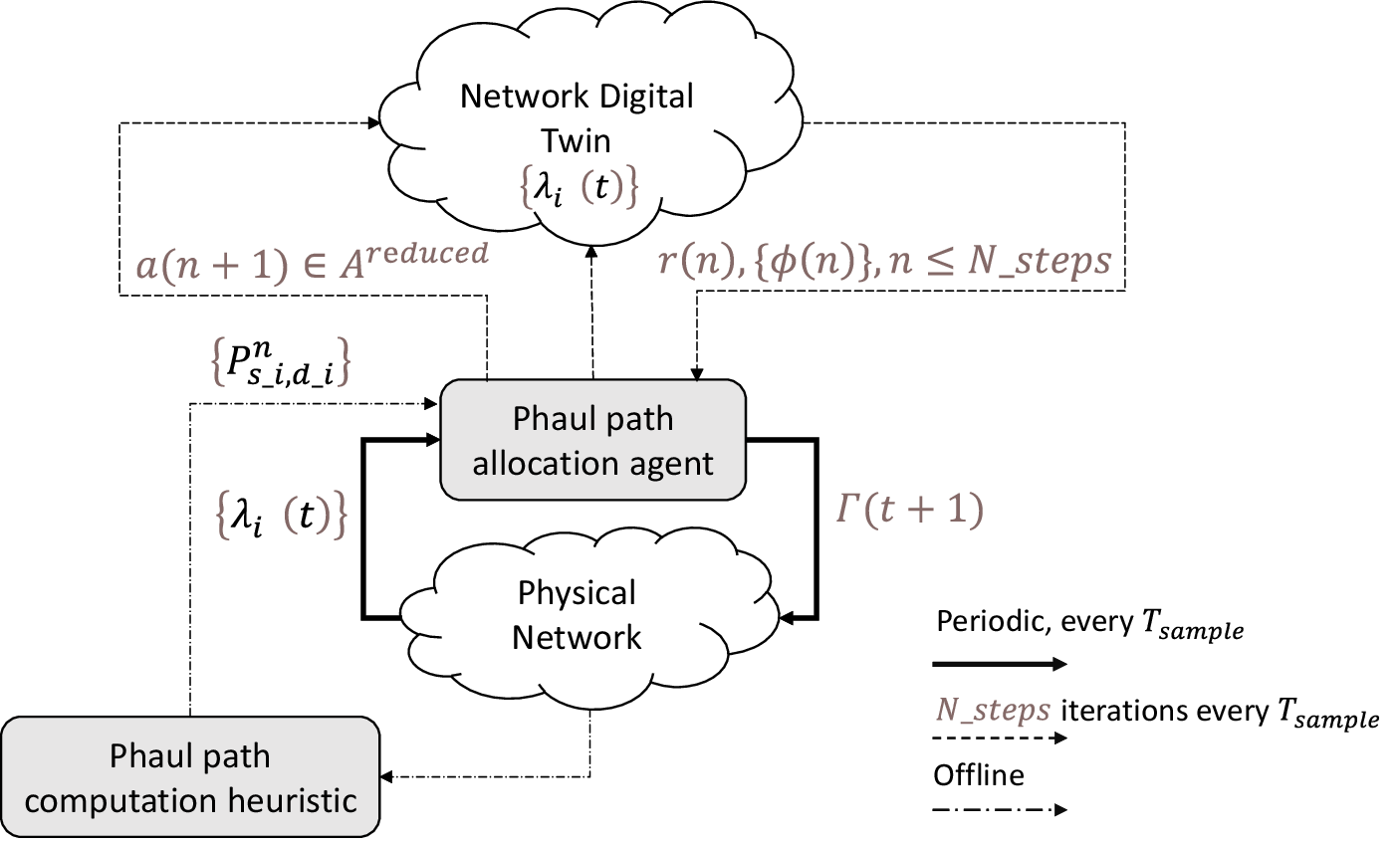}
  \caption{PHaul agent design}
  \label{fig:agent_design}
\end{figure}

Figure \ref{fig:agent_design} describes the high level operation of PHaul. First, we distinguish the two modules that compose PHaul, namely: i) the \emph{path computation heuristic}, and ii) the \emph{path allocation agent}. The path computation heuristic operates offline, and its goal is to examine the topology of the physical network to derive a set of up to $K^{max}$ paths for each backhaul flow through both the Sub6 and the mm-wave backhaul networks. We refer to this set of potential paths as $\mathbb{P}^{selected}_{i}$. The obtained set of paths is only recomputed when the topology of the physical network changes.

The path allocation agent operates online and maintains two loops, one with the physical network and another one with the network digital twin. The agent samples the physical network every $T_{sample}$ to obtain a traffic matrix $\{ \lambda_{i}(t) \}$. Subsequently, for each new sampled traffic matrix, the path allocation engine initializes the digital twin of the wireless backhaul network, and performs up to $N^{steps}$ interactions with the network digital twin to obtain the final flow allocation that will be programmed on the physical network, i.e. $\Gamma(t+1)$. For each interaction with the digital twin, the path allocation agent applies an action that is sampled from a reduced action space, $a(n) \in \mathbb{A}^{reduced}$, where $n$ denotes the iteration with the digital twin. Each action results in an updated path allocation for a given flow that is stored in a vector $\Gamma(n)$, and in a corresponding reward $r(n)$. To be able to compute the reward for a given allocation, the digital twin of the backhaul network models the network topology, the traffic matrix, the flow allocation, and derives an estimate of the effective data rate that would be obtained by each flow under this allocation, i.e. $\phi_{i}(n)$. This process is repeated for $N^{steps}$ or until a termination condition is reached.  
We note that $N^{steps}$ is a key parameter trading off the quality of the obtained allocations with the resulting execution time. This trade-off is studied in detail in Section \ref{sec:perf_eval}.

\subsection{PHaul path computation heuristics}
\label{subsec:path_comp_heur}
The goal of the path computation heuristics is to compute for each backhaul flow $f_{i}$, originating at IAB node $s_{i}$, a set of up to $K^{max}$ potential paths through the Sub6 backhaul network and a set of up to $K^{max}$ potential paths through the mm-wave backhaul network, where paths can be defined towards different IAB donors $d_{k} \in \mathbb{D}$. The output of the path selection heuristic is the set $\mathbb{P}^{selected}_{i}$, defined for each flow $f_{i}$. Notice that the computation of $\mathbb{P}^{selected}_{i}$ is performed offline, using only information about the topology of the backhaul network. 

In this section we describe three path selection heuristics that are applied separately at the Sub6 and the mm-wave backhaul networks. We refer to these heuristics as the \emph{ShortestPath}, the \emph{LastHop} and the \emph{LeastCommon} heuristics. These heuristics are described in Algorithm \ref{alg:path_selection}. 

Before describing each heuristic in detail, we note that for each flow $\mathbb{P}^{all}_{i}$ contains a set of initial potential paths, with $|\mathbb{P}^{all}_{i}| > K^{max}$. The candidate flows in $\mathbb{P}^{all}_{i}$ can be computed using standard path computation algorithms between the source and destinations of each flow. The goal of the path selection heuristics is to select a subset of $K^{max}$ paths from the paths available in $\mathbb{P}^{all}_{i}$.   
Considering that $\mathbb{L}$ indicates the set of links in the network, with $|\mathbb{L}|=M$, we define a path $P_{x}$ as a horizontal vector $P_{x} = [ p_{1}, ..., p_{M}]$ with $p_{j}=1$ if link $j$ belongs to path $P_{x}$, and $p_{j}=0$ otherwise. 

Looking at Algorithm \ref{alg:path_selection}, we can see that \emph{ShortestPath} simply selects for each flow the first $K^{max}$ shortest paths in $\mathbb{P}^{all}_{i}$, where paths $\mathbb{P}^{all}_{i}$ are already sorted in order of hops. \textcolor{black}{Thus, \emph{ShortestPath} has complexity $O(|\mathbb{F}|)$.}

The idea behind \emph{LastHop} is to select for a given flow paths that use a different last hop. The intuition behind this heuristic is that the last hop of a path is going to be reused by many flows. Therefore, allowing flows to use paths with different last hops can enhance the load balancing properties of the network. \textcolor{black}{The complexity of \emph{LastHop} is $O(|\mathbb{F}|K^{max})$, since for each flow all potential paths are evaluated.}

The \emph{LeastCommon} heuristic generalizes the idea behind \emph{LastHop} to all links of a path. To that end, it defines a metric to compare the similarity between two paths by counting the number of common links in each path. This metric can be computed as a dot product between two path vectors, as indicated in line 32 of Algorithm \ref{alg:path_selection}. Thus, \emph{LeastCommon} selects for a given flow the paths that result in the least amount of common links with respect to the already allocated paths for all the other flows. Notice that \emph{LeastCommon} is a greedy heuristic and therefore the final path selection will depend on the order used to explore the set of flows. In our implementation \emph{LeastCommon} starts allocating flows according to their distance from the IAB donor nodes. \textcolor{black}{\emph{LeastCommon} is the most complex heuristic with a complexity of $O(|\mathbb{F}|(\sum_{i=1}^{|\mathbb{F}|}K^{max}*i*|P_{i}^{all}|))$, since each flow is compared to all previously allocated flows. Notice though, that path computation heuristics need only to be run offline, and thus their complexity is not a major concern in practical networks.}

Once the set $\mathbb{P}^{selected}_{i}$ is obtained for each flow, this set is delivered to the path allocation agent, which will consider the current traffic matrix to select the best path for each flow. 

\begin{algorithm}
\caption{PHaul path computation heuristics.}
\label{alg:path_selection}
\SetAlgoLined\DontPrintSemicolon
\SetKwFunction{ShortestPath}{ShortestPath}
\SetKwFunction{LastHop}{LastHop}
\SetKwFunction{LeastCommon}{LeastCommon}

\KwIn{$\mathbb{F}$ set of all flows}
\KwIn{$K^{max}$ number of paths to select in the Sub6 and mm-wave topologies}
\KwIn{$\mathbb{P}^{all}_{i}$ set of all potential paths for flow $f_{i}$ sorted in growing number of hops}
\KwOut{$\mathbb{P}^{selected}_{i}$ set of selected paths for flow $f_{i}$}

\SetKwProg{myproc}{Procedure}{}{}
    \myproc{\ShortestPath{}}{
         \For{$f_{i} \in \mathbb{F}$}{
            $\mathbb{P}^{selected}_{i} \gets \mathbb{P}^{all}_{i}[1:K^{max}]$
         }
    }

\SetKwProg{myproc}{Procedure}{}{}
    \myproc{\LastHop{}}{
     $\mathbb{L}^{last} \gets \emptyset$ \\
        \For{$f_{i} \in \mathbb{F}$}{
                       \For{$P_{x} \in \mathbb{P}^{all}_{i}$}{
                $last\_link \gets$ last link in $P_{x}$ \\
                \If{last\_link $\notin \mathbb{L}^{last}$}{
                    $\mathbb{L}^{last} \gets Add(last\_link)$
                    $\mathbb{P}^{selected}_{i} \gets Add(P_{x})$                
                }
                \If{$|\mathbb{P}^{selected}_{i}|=K^{max}$}{
                    go to next flow
                }
            }
            \For{$P_{x} \in \mathbb{P}^{all}_{i}$}{
                \If{$P_{x} \notin \mathbb{P}^{selected}_{i}$}{
                    $\mathbb{P}^{selected}_{i} \gets Add(P_{x})$   
                }
                \If{$|\mathbb{P}^{selected}_{i}|=K^{max}$}{
                    go to next flow
                }                
            }
        }       
    }

\SetKwProg{myproc}{Procedure}{}{}
    \myproc{\LeastCommon{}}{
        $\mathbb{P}^{selected}_{ALL} \gets \emptyset$ \\
        \For{$f_{i} \in \mathbb{F}$}{
            \For{$P_{x} \in \mathbb{P}^{all}_{i}$}{
                $comn\_links \gets 0$ \\
                \For{$P_{y} \in \mathbb{P}^{selected}_{ALL}$}{
                    $comn\_links \gets comn\_links + P_{x}P_{y}^{T}$
                }
                $\mathbb{P}^{unsorted}_{i} \gets Add(P_{x}, comn\_links)$ \\
            }
            $\mathbb{P}^{sorted}_{i} \gets SortIncreasing(\mathbb{P}^{unsorted}_{i})$ \\
            $\mathbb{P}^{selected}_{i} \gets \mathbb{P}^{sorted}_{i}[1:K^{max}]$\\
            $\mathbb{P}^{selected}_{ALL} \gets Add(\mathbb{P}^{selected}_{i})$\\
        }
    }

\end{algorithm}

\subsection{PHaul path allocation agent}
\label{subsec:path_alloc_agent}
Algorithm \ref{alg:phaul} depicts the detailed operation of the PHaul path allocation agent, where the following variables are considered:
\begin{itemize}
\item[i.] $NetDTwin$ is a digital twin of the wireless backhaul network, which allows to perform flow allocations and to compute the resulting per-flow effective data rates, $\phi_{i}(n)$, and the resulting rewards $r(n)$, (see Figure \ref{fig:agent_design}).
\item[ii.] The set of paths available to $f_{i}$ defined as a vector of size $2K^{max}$, with the following components $\mathbb{P}^{selected}_{i} = \{ P^{sub6}_{i,1}, P^{mwv}_{i,1}, ..., P^{sub6}_{i,K^{max}}, P^{mwv}_{i,K^{max}}\}$, where $P^{j}_{sub6}$ and $P^{j}_{mwv}$ represent respectively the $j-th$ path for this flow in the Sub6 or mm-wave backhaul networks. 
\item[iii.] The allocation vector $\Gamma_{i}$, $i \leq |\mathbb{F}|$, where $|\mathbb{F}|$ is the number of flows, and the $i-th$ component of this vector contains the index within $\mathbb{P}^{selected}_{i}$ representing the path allocated to flow $f_{i}$. 
\end{itemize}
Based on the previous variables we describe now the action space, the state space and the reward used by the PHaul path allocation agent.

\subsubsection{Action Space} It is defined as an integer $0 \leq a \leq 2K^{max}|\mathbb{F}|$, where $|\mathbb{F}|$ is the number of flows and $K^{max}$ the number of paths available per flow in the Sub6 and mm-wave networks. Each action $a$ encodes a single flow allocation in the following way:
\begin{itemize}
    \item[i.] Action $a$ represents flow $f_{i} = a \mod{|\mathbb{F}|}$
    \item[ii.] $\Gamma_{i} = \lfloor \frac{a}{|\mathbb{F}|} \rfloor$ indicates the index within $\mathbb{P}^{selected}_{i}$ corresponding to the path allocated to $f_{i}$ (cf. Algorithm \ref{alg:phaul} line 22).
\end{itemize}

Thus, we can see that selecting action $a$ represents allocating to flow $f_{i} = a \mod{|\mathbb{F}|}$ its path number $\lfloor \frac{a}{|\mathbb{F}|} \rfloor$ within the vector $\mathbb{P}^{selected}_{i}$. Repeated iterations of the policy will result in allocating paths for different flows.

\subsubsection{State Space} Consider $\lambda_{i}(t)$ the input data-rate for flow $f_{i}$ measured at its source $s_{i} \in \mathbb{S}$ in the last sampling interval, and $\phi_{i}(n)$, with $n < N^{steps}$, the effective data-rate for this flow measured at its destination IAB donor $d_{i} \in \mathbb{D}$ in the last interaction with the network digital twin\footnote{Notice that $\phi_{i}(n)$ depends on the allocated path and is updated at each interaction of the PHaul agent with the network digital twin, whereas $\lambda_{i}(t)$ is only updated upon sampling the physical network (c.f. Figure \ref{fig:agent_design})}. Then, the environment state in PHaul is modeled as a vector containing the collection of input and effective data-rates for each flow, as well as the current path allocated to each flow (cf. Algorithm \ref{alg:phaul} line 20).

\textcolor{black}{Notice that the state space in PHaul requires only the total number of IAB nodes generating backhaul flows, but it does not require detailed internal knowledge of the topology, i.e. nodes and edges.}  This is a design decision to make PHaul robust to \textcolor{black}{internal} changes on the topology, which can occur for example if a backhaul link is blocked. The PHaul state space captures the input and output traffic matrices, as well as the current paths allocated to each flow. Thus, the intuition behind the PHaul path allocation agent is that it should learn to correlate types of traffic matrixes to the specific paths that result in a good allocation for that traffic matrix, while considering the IAB network to be a black box. \textcolor{black}{In Section \ref{sec:perf_eval} we present experimental results that validate this hypothesis.}

\subsubsection{Reward definition} The PHaul reward is modelled after the traffic engineering criteria defined by the network operator. For example, in the case of the weighted traffic engineering criteria defined in Section \ref{sec:network_model}, the reward $r(n)$, $n < N^{steps}$, is defined for each interaction of the agent as:  
\begin{equation}
    r(n) = (1 + \gamma) \frac{\sum_{i} \phi_{i}(n)}{\sum_{i} \lambda_{i}(t)} + (1 - \gamma)\frac{(\sum_{i} \phi_{i}(n))^{2}}{|\mathbb{F}|\sum_{i} \phi_{i}(n)^{2}}
\end{equation}
Where $t$ is the last time that the traffic matrix was sampled from the physical network. We can see that $r(n)$ encodes a balance between throughput and fairness regulated by the parameter $\gamma$. Let us therefore define the reward as:
\begin{equation}
    r(n) = (1 + \gamma) thr(n) + (1 - \gamma) fair(n)
\end{equation}
Now, given that we cannot know a priory how far from $1$ both $thr(n)$ and $fair(n)$ will be in a practical network, we redefine the reward as:

\begin{equation}
    -2\leq \hat{r}(n) = ( 1 + \gamma) \hat{thr}(n) + (1 - \gamma) \hat{fair}(n) \leq2
\end{equation}
\begin{equation}
\label{eq:thr_fair_norm}
        \hat{J}(n) = \frac{(J(n) - J_{min}) - (J_{max} - J(n))}{J_{max} - J_{min}}
\end{equation}
    
Where $\hat{thr}(n)$ and $\hat{fair}(n)$ are computed according to equation \ref{eq:thr_fair_norm}, and are constantly updated based on the rewards obtained in each interaction with the network digital twin (cf. Algorithm \ref{alg:phaul} line 3). Note that normalizing the reward to the maximum and minimum throughput or fairness values obtained throughout the $N^{steps}$ iterations with the network digital twin is helpful for the agent to understand if the actions being applied across the different iterations are pushing the reward in the right direction. 

\subsubsection{Termination condition} We need at least to execute the allocation policy $N^{steps}=|\mathbb{F}|$ iterations to have the opportunity to allocate a path for each flow, and up to $N^{steps}=(2K^{max})^{|\mathbb{F}|}$ iterations, to visit all potential allocations. In practice, $N^{steps}$ is a hyper-parameter that allows to trade-off accuracy and inference time. Section \ref{sec:perf_eval} evaluates the impact of this parameter on training accuracy. The PHaul agent terminates after executing $N^{steps}$ iterations over the network digital twin, or as soon as an allocation is found that results in an optimal reward (cf. Algorithm \ref{alg:phaul} line 18).

\subsubsection{Selected DRL algorithm}
Based on the previous definitions of action space, state space and reward, any DRL agent able to operate with a discrete action space could be applied to PHaul. In this work, we base our implementation on the Proximal Policy Optimization (PPO)~\cite{ppo} algorithm, which achieves state-of-the-art performance across a wide range of challenging tasks and outperforms several other DRL algorithms~\cite{ppo_comp}. We leave as future work the study of additional DRL policies to PHaul. 

\begin{algorithm}[ht]
  \SetAlgoLined\DontPrintSemicolon
  \SetKwFunction{PHaul}{PHaul}
  \SetKwFunction{ComputeReward}{ComputeReward}
  \SetKwFunction{Init}{Init}
  \SetKwProg{myalg}{Algorithm}{}{}

    \SetKwProg{myproc}{Procedure}{}{}
      \myproc{\ComputeReward{}}{
      $\phi_{i}(n) = NetDTwin.EffectiveRates()$\;
      Compute $\hat{thr}(n)$, $\hat{fair}(n)$\;
      $\hat{r}(n) = ( 1 + \gamma) \hat{thr}(n) + (1 - \gamma)\hat{fair}(n)$\;
      $thr_{max} = \max \{ thr(n), thr_{max}\}$\;
      $thr_{min} = \min \{ thr(n), thr_{min}\}$\;
      $fair_{max} = \max \{ fair(n), fair_{max}\}$\;
      $fair_{min} = \min \{ fair(n), fair_{min}\}$\;
    }
    
    \SetKwProg{myproc}{Procedure}{}{}
      \myproc{\Init{}}{ 
      $n=0$\;
      $NetDTwin.Init(\{\lambda_{i}(t) \})$\;
      $\Gamma_{i} = 0$, $\forall i \in \mathbb{F}$\;
      $thr_{max} = fair_{max}=-\infty$\;
      $thr_{min} = fair_{min}=\infty$\;
      $NetDTwin.Alloc(\Gamma(n))$\;
    }
    
    \myalg{\PHaul{}}{
      \KwIn{$\{ \lambda_{i}(t) \}$, $NetDTwin$, $|\mathbb{F}|$, $\gamma$}
      \KwOut{$\Gamma$ vector with per-flow allocations}
      \Init{}\;
      \While{$n < N^{steps}$ or $r(n) <2$}{
            $\phi_{i}(n) = NetDTwin.EffectiveRates()$\;
            $state = \{ {\lambda_{i}(t)}, {\phi_{i}(n)}, \Gamma (n)\}$\;
            $a = PPO.action()$\;
            $\Gamma(a \mod{|\mathbb{F}|}) = \lfloor \frac{a}{|\mathbb{F}|} \rfloor$\;
            $NetDTwin.Alloc(\Gamma(n))$\;        
            \ComputeReward{}\;
            $n = n+1$
        }        
    }

\caption{PHaul agent execution in inference mode}
\label{alg:phaul}
\end{algorithm}

\subsection{Visualizing a PHaul path allocation}
\label{subsec:visual_allocation}

\begin{figure*}[t]
  \centering
  \includegraphics[width=0.45\linewidth]{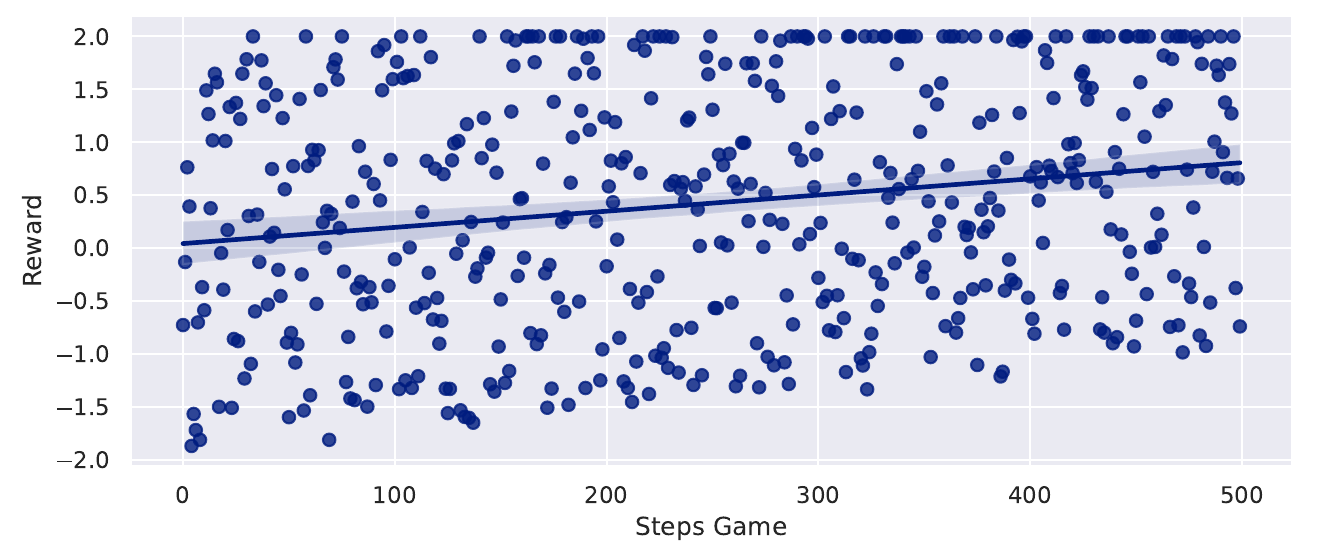}
  \includegraphics[width=0.45\linewidth]{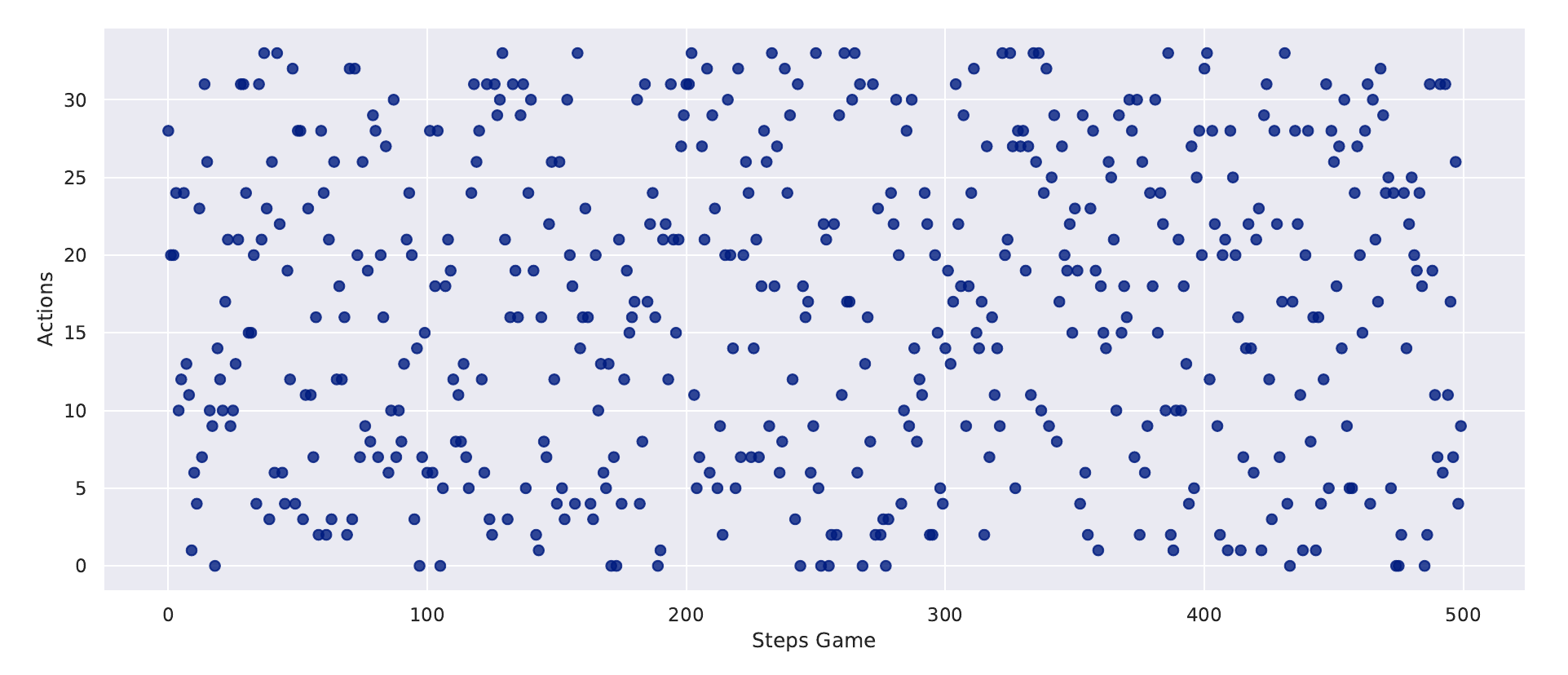}
  \caption{Visualizing a PHaul path allocation. Left, reward $\hat{r}(n)$ across iterations with the digital twin. Right, action $a(n)$ across iterations with the digital twin.}
  \label{fig:visualizing_phaul}
\end{figure*}

To visualize the operation of the PHaul path allocation agent, Figure \ref{fig:visualizing_phaul} depicts, on the left, the instantaneous value of the normalized reward $\hat{r}(n)$ obtained by a trained PHaul agent across $N^{steps}=500$ iterations, and, on the right, the instantaneous action $a(n)$ being selected, where each action represents the allocation for an individual flow.

The setup considered in Figure \ref{fig:visualizing_phaul} consists of a network of 17 IAB nodes and three IAB donor nodes, where each IAB node generates a backhaul flow, and $K^{max}=1$ paths per flow are considered in the Sub6 and mm-wave backhaul networks. This results in a reduced action space of $2K^{max}{|\mathbb{F}|}=34$ actions. 

We see in the left part of Figure \ref{fig:visualizing_phaul} the instantaneous normalized reward $\hat{r}(n)$, as well as a regression of $\hat{r}(n)$, clearly showing how the agent becomes more efficient as the number of iterations increases, where we see a saturated $\hat{r}(n)=2$ when $n$ approaches $N^{steps}$. However, recall from Algorithm \ref{alg:phaul}, that PHaul does not necessarily return the allocation explored in the last iteration, but instead it will return the best performing allocation sampled across the $N^{steps}$ iterations. It is clear from the left part of Figure \ref{fig:visualizing_phaul} that $N^{steps}$ is a critical parameter in the performance of PHaul, and as such it will be studied in detail in Section \ref{sec:perf_eval}.

Looking now at the right part of Figure \ref{fig:visualizing_phaul}, which depicts the action $a(n)$ sampled by PHaul across the iterations with the digital twin, we can see how indeed PHaul samples potential allocations for all flows. Notice that PHaul often returns to the same action, as the effect of an action will differ depending on the current allocation for all the other flows. This is why PHaul includes the current allocation as part of the state space definition.

\subsection{\textcolor{black}{Implementation discussion}}
\label{subsec:implementation}
\textcolor{black}{To deploy PHaul in a practical 3GPP IAB system it is required to have an interface that allows the PHaul path allocation agent to dynamically update the BAP path to be used for each backhaul flow. This interface should allow to bind a backhaul flow identifier, which could be based for example on the GTP Tunnel EndPoint ID, with the BAP path identifier, which is already included in the BAP header. This interface can be implemented in the context of the O-RAN architecture by defining an E2 service model tailored to IAB networks, as already proposed in \cite{iab-oran}. Thus, with an IAB E2 service model  being available, the PHaul path allocation agent could be deployed as an xApp in the near real-time RAN Intelligent Controller (RIC), whereas the path computation heuristic and the PHaul training could be implemented as part of the non-real time RIC. Additionally, A1 policies could be used to specify the parameter $\gamma$ in the xApp according to the desired traffic engineering criteria.}

\section{Performance Evaluation}
\label{sec:perf_eval}
We present in this section an in-depth performance evaluation of PHaul that is organized in the following way. In section \ref{subsec:eval_setup} we describe our simulation setup and modelling assumptions. In section \ref{subsec:comp_heuristics} we describe the competing heuristics that we use to benchmark PHaul. In section \ref{subsec:pe_path_comp} we evaluate the PHaul path computation heuristics. In section \ref{subsec:pe_path_alloc_training} we optimize the training hyper-parameters of the PHaul path allocation agent, which is then benchmarked against competing approaches in section \ref{subsec:pe_path_alloc_inference}. Finally, in sections \ref{subsec:pe_path_alloc_exec_time}, \ref{subsec:pe_path_alloc_broken}, and \ref{subsec:pe_path_alloc_untrained_topos} we look respectively at the execution time required by the PHaul path allocation agent, \textcolor{black}{at the impact of broken links, and at the ability of this agent to generalize to unseen backhaul topologies plus the gains that are due to the addition of Sub6 capabilities to the IAB backhaul.}

\subsection{Evaluation Setup}
\label{subsec:eval_setup}

To evaluate PHaul we have developed a flow-level simulation model in Python, connected to OpenAI's Gym~\cite{openai_gym} environment, which is used to implement the PHaul path allocation agent based on PPO. We leverage the PPO implementation available in the \emph{stable-baselines} package\footnote{\url{https://stable-baselines3.readthedocs.io/en/master/}}. The developed simulator can be understood as the network digital twin component described in Section \ref{sec:agent_design}, i.e. it allows to generate random backhaul topologies and traffic matrices, to allocate a backhaul flow through a given path, and to estimate the effective capacity, $\phi_{i}$, obtained by a flow under a given set of flow allocations.

To model representative IAB backhaul topologies, we developed a random topology generator after the exemplary IAB network reported in \cite{iab-deployment}, which covers a suburban area in Chicago, US. In \cite{iab-deployment}, the authors study different types of IAB deployments in a network consisting of up to 45 IAB nodes, where the number of donor nodes varies between 8 and 30 according to the fiber penetration in the area. Depending on the number of donors considered, the authors report chains of up to 3 hops and a mean hop count between 1.5, when the number of donors is large, and 2.08, when the number of donors is small. For our study, we are interested in scenarios where fibre penetration is low, thus we consider IAB topologies with a small number of donors, and study how these networks perform as we increase the number of non-donor IAB nodes. 

We generate IAB topologies in the following way. First, we start with a first layer of 3 donor nodes. Then, for each donor node we generate $n$ IAB child nodes, with $n$ being a uniform random variable between 1 and 3. We repeat this process for each layer of IAB nodes until the target number of nodes in the topology is reached. All IAB nodes have a wireless link to their parent node, but we also allow for nodes to have multiple parents using a random parameter, \emph{edge\_selection\_prob}, that adds an additional link between an IAB node and the IAB node located next to its parent in the upper layer. This parameter allows us to control the presence of multiple paths from a given IAB node to the wired network. Notice, that while being random, this topology generation process reflects the topology formation process in a real IAB network that would start with a fixed number of donor nodes, and then progressively grow to expand the network footprint. 

Following our network model in Figure \ref{fig:network_model}, the same IAB nodes are available in the Sub6 and mm-wave topologies, but the exact links between nodes may differ in each topology. In particular, we set \emph{edge\_selection\_prob} to $0.4$ in the mm-wave topology and to $0.6$ in the Sub6 topology to reflect the fact that Sub6 links have wider coverage. Following this approach we consider scenarios with a total number of IAB nodes (including donor nodes) varying from 20 to 60, which result in mean hop counts of $3.08$ for the case of 20 nodes and $5.12$ for the case of 60 nodes.

As described in Section \ref{sec:network_model}, we model backhaul links as interference-free with capacities that are stable during the execution of the agent. In the case of the mm-wave backhaul, we assume a 30 GHz deployment with a 800 MHz carrier bandwidth, resulting in backhaul downlink per-link capacities in suburban environments that we select randomly between 800 Mbps and 1 Gbps, according to the simulation results reported in \cite{ericsson-iab}. For the case of Sub6 links we consider an interference-free 80 MHz link with capacities that we select randomly between 200 and 300 Mbps, according to the Sub6 wireless backhaul measurements reported in \cite{5gxhaul-bristol}.

Regarding the input traffic matrix, we consider that each IAB node aggregates the traffic from all connected UEs into a single IAB backhaul flow. Thus, the traffic carried by this flow will depend on the simultaneous number of UEs in RRC\_CONNECTED state in each IAB node, and on the capacity of the access interface. Lacking this type of data from real deployments, we model the load offered by each backhaul flow in the following way. First, we use a random parameter, \emph{node\_active\_probability}, to determine if there are active UEs, i.e. in RRC\_CONNECTED mode, in that node. In case there are active UEs, we model the resulting backhaul traffic as a uniform random variable between $\lambda_{min}$ and $\lambda_{max}$, where we consider two scenarios with a growing flow size in Mbps, namely: i) $\lambda_{min}=250$, $\lambda_{max}=500$, and ii) $\lambda_{min}=500$, $\lambda_{max}=750$.

Finally, to achieve statistically significant results, every time we report a performance figure for a given network configuration, we consider at least 10 random topologies for that network configuration, and for each of those topologies we average the results of 250 randomized input traffic matrixes. Unless otherwise stated, the PHaul path allocation agent is trained for each specific topology considered. The metrics reported in this section correspond to average values that are depicted with their corresponding 95\% confidence interval, which is however too small to be clearly seen in the figures. 

\textcolor{black}{To ease the reproducibility of the results presented in this section, we have released as open source our implementation of PHaul, the implementation of the IAB network digital twin, and all the supporting evaluation environments\footnote{\textcolor{black}{https://github.com/Fundacio-i2CAT/phaul/}}.}

\subsection{Competing heuristics}
\label{subsec:comp_heuristics}
To assess the performance of PHaul, we focus on two metrics: i) the score achieved in terms of the weighted objective function $J^{TE}$ introduced in Section \ref{sec:network_model}, and ii) the required execution time. Recall that the weighted objective function is defined as $J^{TE} = (1 + \gamma) \frac{\sum_{i} \phi_{i}(t)}{\sum_{i} \lambda_{i}(t)} + (1 - \gamma)\frac{(\sum_{i} \phi_{i}(t))^{2}}{|\mathbb{F}|\sum_{i} \phi_{i}(t)^{2}}$, where we can set $\gamma=1$ to focus exclusively on throughput efficiency, hereafter referred to simply as efficiency, and $\gamma=-1$ to focus on inter-flow fairness.

PHaul is then compared to three alternative path allocation agents, namely: i) \emph{brute force}, ii) \emph{subset-sum}, and iii) \emph{random}. All the evaluated path allocation agents use the same path selection heuristic (c.f. Section \ref{sec:agent_design}).
 
The \emph{brute force} path allocation agent considers, in the digital twin, all combinations between flows and paths and selects the one yielding the highest reward $\hat{r}(n)$. Thus, for a given number of flows $|\mathbb{F}|$ and paths $K^{max}$, brute force explores $(2K^{max})^{|\mathbb{F}|}$ possible allocations in the network digital twin. For every considered allocation, brute force evaluates the objective function $J^{TE}$, with $\gamma$ set accordingly to maximize efficiency or fairness. Brute force is optimal in terms of the objective function, but it leads to large execution times and it can only be used for network configurations with a limited number of flows and paths.
    
The \emph{subset-sum} path allocation agent is a greedy heuristic based on the subset-sum problem~\cite{subset-sum}. Subset-sum orders backhaul flows in decreasing order of their traffic demand $\lambda_i$, and allocates each flow sequentially. For each flow, the agent explores allocations in the $2K^{max}$ paths available to that flow, and selects the one that maximizes the objective function $J^{TE}$, with $\gamma$ set accordingly. Subset-sum scales linearly with the number of flows $|\mathbb{F}|$, evaluating up to $2K^{max}|\mathbb{F}|$ allocations.

Finally, the \emph{random} path allocation agent, simply selects for each flow one of the $2K^{max}$ paths available through the mm-wave and Sub6 backhaul networks using a uniform random variable, scaling linearly with the number of flows $|\mathbb{F}|$.

\subsection{Evaluating the PHaul path computation heuristics}
\label{subsec:pe_path_comp}

\begin{figure*}[h]
    \centering

    \subfigure[Efficiency ($\gamma=1$) for $K^{max}=1$]{
        \includegraphics[width=0.48\textwidth]{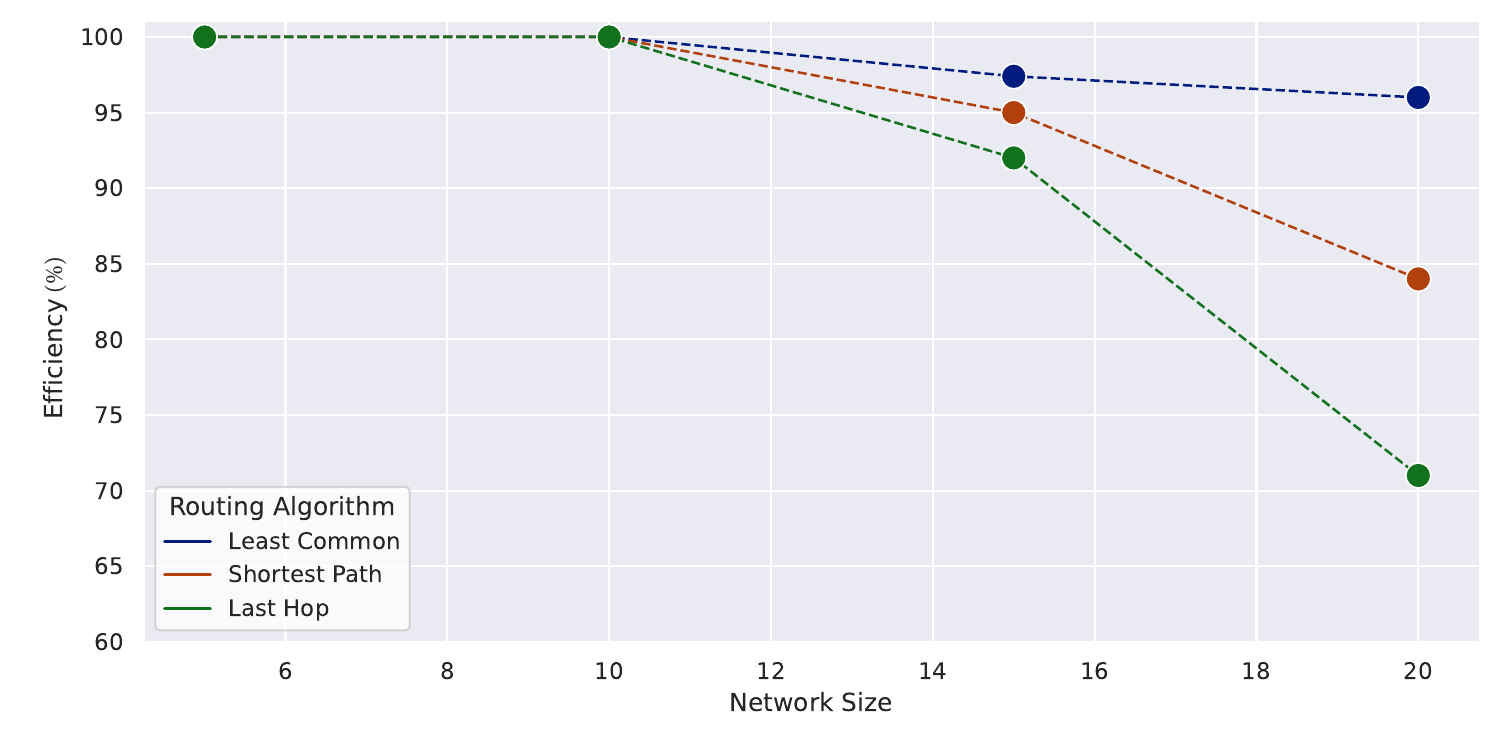}
        \label{fig:subfig1}
    }
    \hfill
    \subfigure[Fairness ($\gamma=-1$) for $K^{max}=1$]{
        \includegraphics[width=0.48\textwidth]{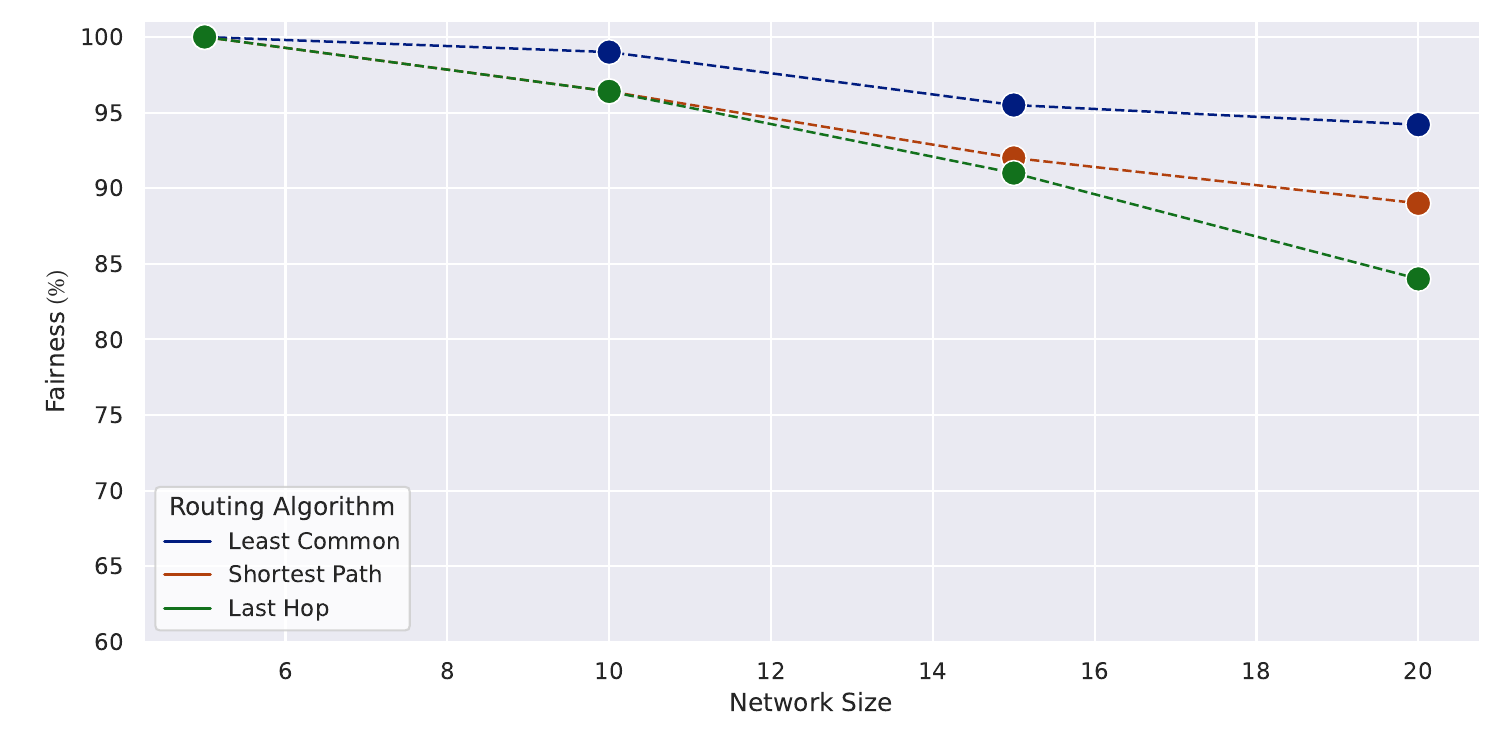}
        \label{fig:subfig2}
    }

    \caption{Impact of PHaul path computation heuristics assuming a brute-force forwarding agent}
    \label{fig:path_comp_heur_eval}
\end{figure*}

We start our evaluation by looking at the path computation heuristics described in Section \ref{subsec:path_comp_heur}. To isolate the effect of the path computation heuristics from the impact of the path allocation agent, we evaluate the different path computation heuristics using the brute force path allocation agent. \textcolor{black}{We note that using the brute force path allocation agent is not possible in practice, but our goal in this experiment is to compare path computation heuristics under a common setting, whithout the influence of the PHaul agent coming into play.} However, using the brute force path allocation agent severely limits the size of the topologies that we can test. Hence, we limit this evaluation to network sizes of up to 20 IAB nodes (including the 3 donor nodes), and only $K^{max}=1$ paths are selected for each flow in the Sub6 and the mm-wave networks. To stress the network, we set \emph{node\_active\_probability} to 1 and consider an input traffic matrix characterized by $\lambda_{min}=500$ Mbps, $\lambda_{max}=750$ Mbps.

Figure \ref{fig:path_comp_heur_eval} depicts the performance of the \emph{ShortestPath} (orange), \emph{LastHop} (green) and \emph{LeastCommon} (blue) path computation heuristics on throughput efficiency (left graph), with $\gamma=1$ in $J^{TE}$, and fairness (right graph), with $\gamma=-1$ in $J^{TE}$. We can see in Figure \ref{fig:path_comp_heur_eval} how the \emph{LeastCommon} heuristic clearly outperforms the \emph{ShortestPath} and \emph{LastHop} heuristics both in terms of throughput efficiency and fairness. The performance of the different path computation heuristics is affected by the IAB topologies considered in our evaluation, which, as previously described, model realistic deployments that tend to follow a tree-like structure that arises naturally when the IAB network is built progressively starting from the donor nodes. Under this type of network topology, flows originating at a given IAB node may share links in their path with all IAB flows that originate at higher levels of the topology. Therefore, heuristics like \emph{ShortestPath} or \emph{LastHop}, which focus only on the path allocations for a single flow at a time, result in links being reused by many flows. On the other hand, \emph{LeastCommon}, by considering the path allocations of previous flows and minimizing the number of joint links across flows, is able to select paths that are more disjoint, resulting in an increased effective capacity in the network.

We hereafter use the \emph{LeastCommon} heuristic as our default path computation heuristic.

\subsection{Training the PHaul path allocation agent}
\label{subsec:pe_path_alloc_training}

We evaluate in this section the training phase of the PHaul path allocation agent. Our goal is twofold. First, we want to understand how PHaul's performance depends on our training hyper-parameters, namely $N^{steps}$ used in Algorithm \ref{alg:phaul}, and the parameter \emph{training\_steps}, which determines the overall number of steps considered in the training phase of PPO \cite{ppo}. Second, we want to benchmark the performance gap between PHaul and the brute force agent, which provides an optimal performance. As performance metrics, we look separately at efficiency obtained with $\gamma=1$ in $J^{TE}$, and then at fairness, obtained with $\gamma=-1$ in $J^{TE}$.

\begin{figure*}[ht]
    \centering

    \subfigure[Impact of $N^{steps}$ on efficiency with $training\_steps=10^{5}$]{
        \includegraphics[width=0.48\textwidth]{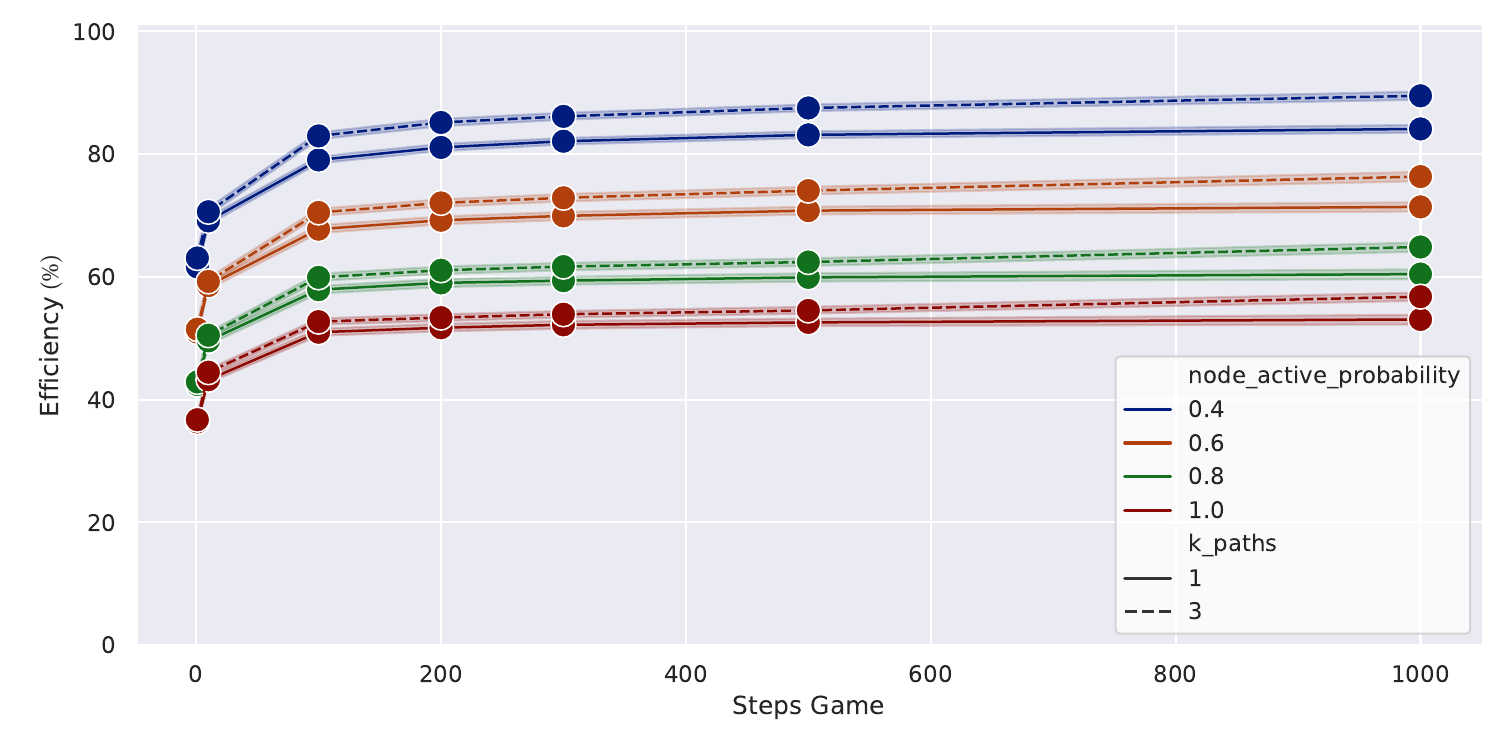}
        \label{fig:subfig1}
    }
    \hfill
    \subfigure[Impact of $N^{steps}$ on fairness with $training\_steps=10^{5}$]{
        \includegraphics[width=0.48\textwidth]{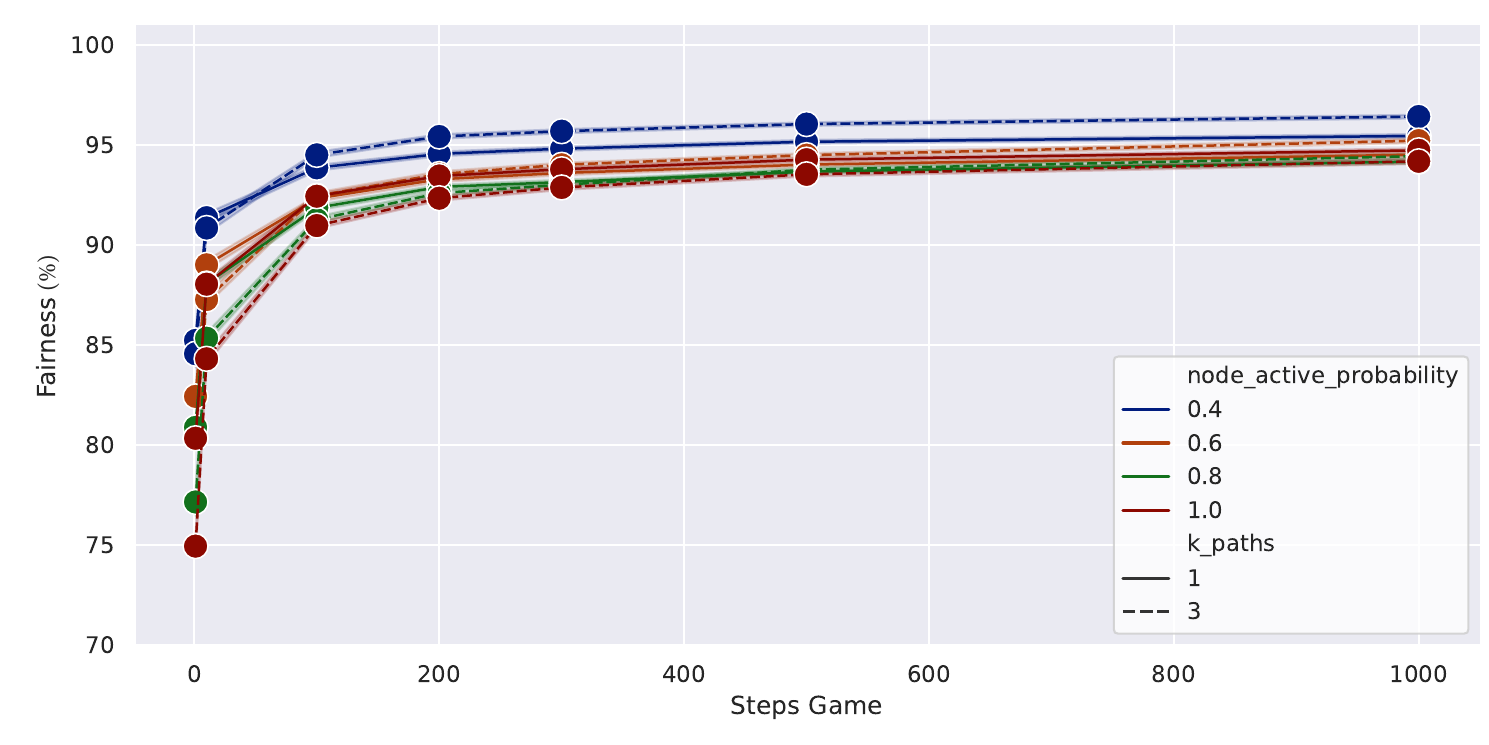}
        \label{fig:subfig2}
    }

    \subfigure[Impact of $training\_steps$ on efficiency with $N^{steps}=300$]{
        \includegraphics[width=0.48\textwidth]{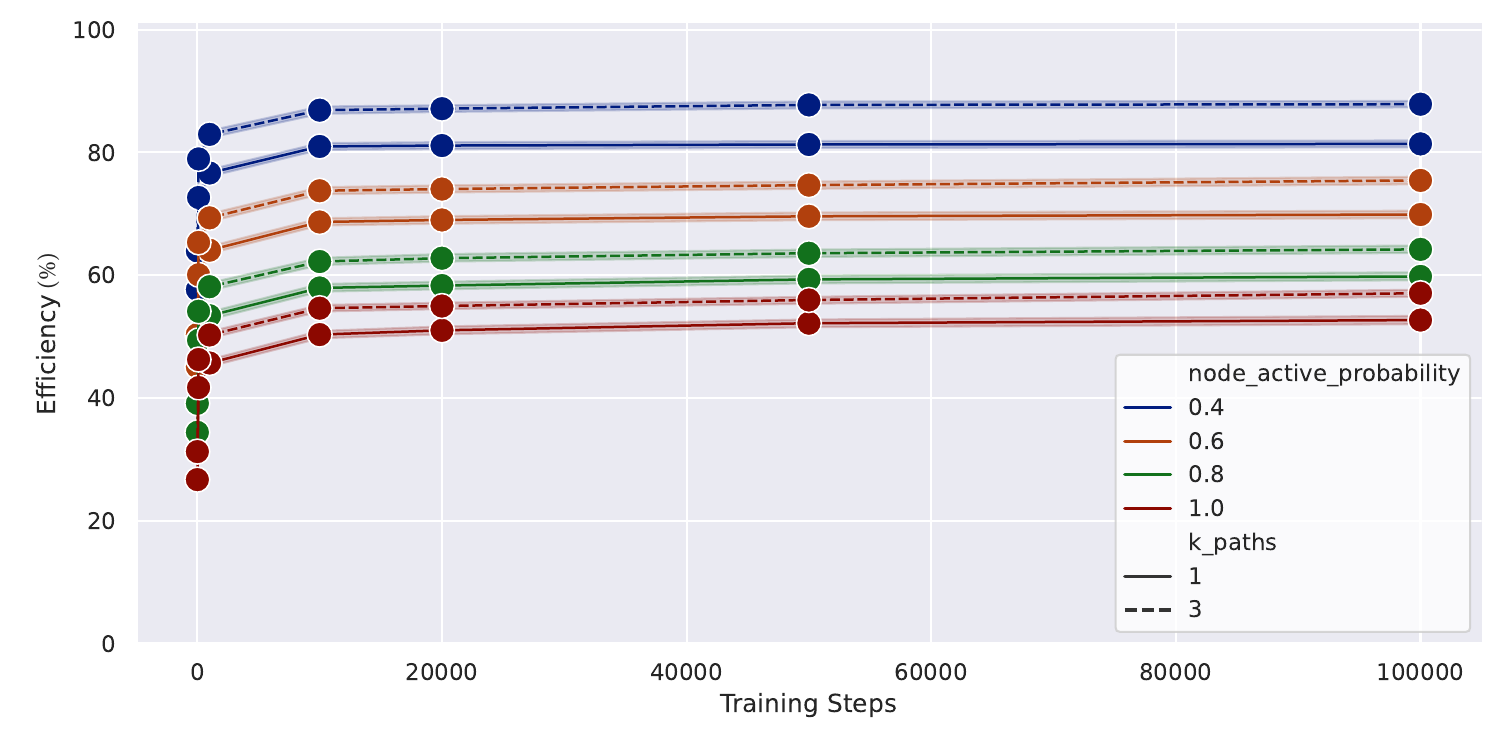}
        \label{fig:subfig3}
    }
    \hfill
    \subfigure[Impact of $training\_steps$ on fairness with $N^{steps}=300$]{
        \includegraphics[width=0.48\textwidth]{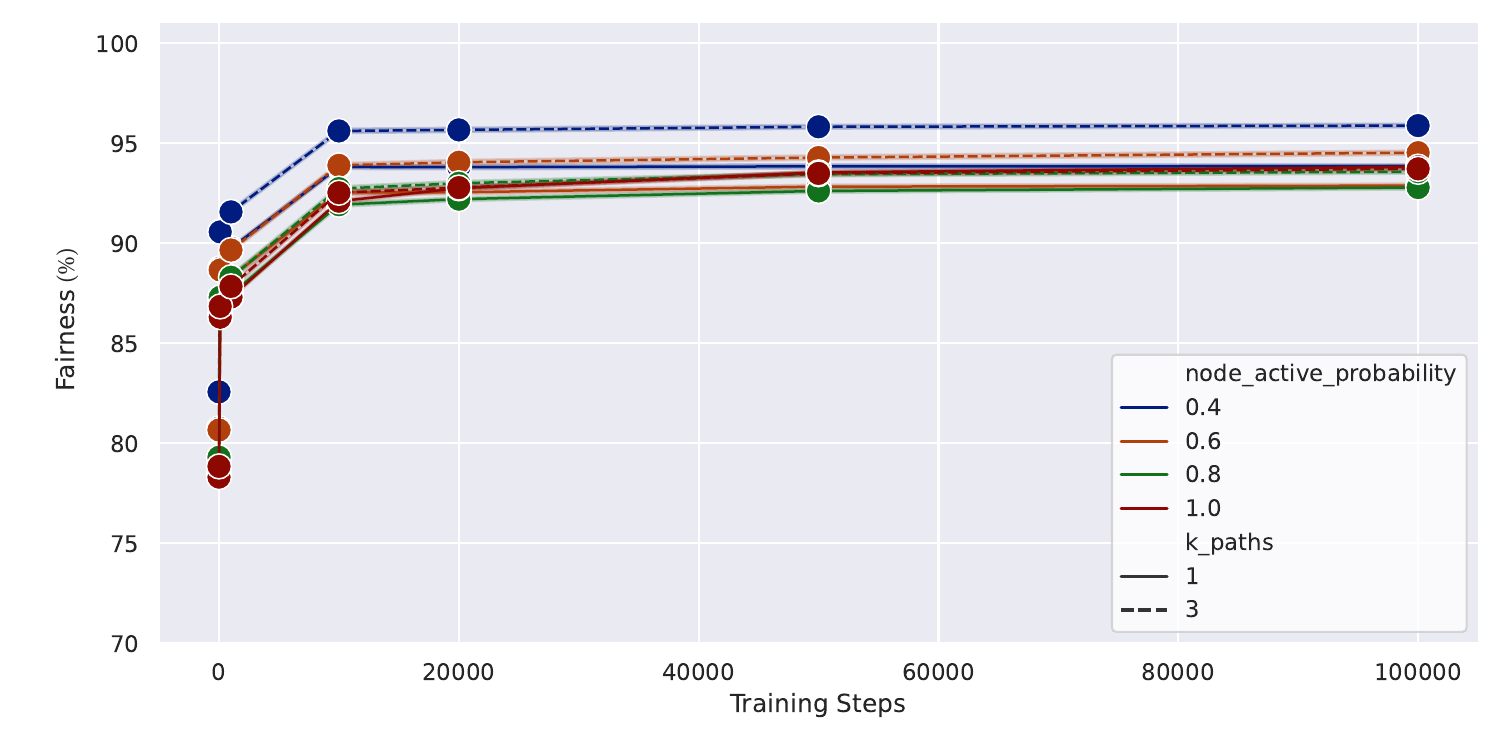}
        \label{fig:subfig4}
    }

    \caption{Training against Brute Force}
    \label{fig:phaul_training_eval}
\end{figure*}

Figure \ref{fig:phaul_training_eval} depicts in the upper row the impact of $N^{steps}$ while fixing \emph{training\_steps} to $10^{5}$, and in the lower row the impact of \emph{training\_steps} while fixing $N^{steps}$ to $300$. In these experiments, we consider an IAB network size of 50 nodes, and vary the \emph{node\_active\_probability} parameter between 0.4 and 1, while considering a random backhaul flow rate between $\lambda_{min}=500$ Mbps, $\lambda_{max}=750$ Mbps. We depict experiments for $K^{max}=1$ and $K^{max}=3$ paths in the Sub6 and mm-wave networks. 

We can see how, both for efficiency and fairness, performance is low when $N^{steps}$ and \emph{training\_steps} are small, and smoothly increases when these parameters grow. Notice though that $N^{steps}$ will have an impact on the execution time of a trained agent, while \emph{training\_steps} only affects the overall training time, which is not a critical parameter. Looking at the knee exhibited by the training curves, we can see that this is independent from the \emph{node\_active\_probability} parameter, which means that regardless of the level of activity in the network the PHaul agent training performance is maintained. The number of paths $K^{max}=1$ and $K^{max}=3$ does not impact either the position of the knee in the training curves. Based on these results the training of PHaul agent can be simplified by setting a fixed value of $N^{steps}$ and \emph{training\_steps}, regardless of the number of paths or the amount of active flows, which relaxes the requirements to train the PHaul network digital twin in real networks. Hereafter we consider $N^{steps} = 300$ and $training\_steps=20000$.

We observe on the left part of Figure \ref{fig:phaul_training_eval} how efficiency decreases when increasing \emph{node\_active\_probability}. This is expected because a higher \emph{node\_active\_probability} means more load being injected into the wireless backhaul, which is saturated in all cases. The impact of \emph{node\_active\_probability} is however not so clear when looking at fairness (right part of Figure \ref{fig:phaul_training_eval}). The reason is that the regardless of the number of backhaul flows active in the network, PHaul is able to allocate the bottleneck bandwidth across these flows in a fair way. Looking at the impact of varying the number of paths $K^{max}$ from 1 to 3, as expected, we observe that considering more paths leads to higher network efficiency as flows can be better balanced through the network. Notice though, that the potential gain achieved by increasing $K^{max}$ depends on the path diversity available in the IAB topology, which is limited in our scenarios that mostly consist in tree-like topologies with limited multi-path opportunities.

\begin{figure}[ht]
  \centering
  \includegraphics[width=\linewidth]{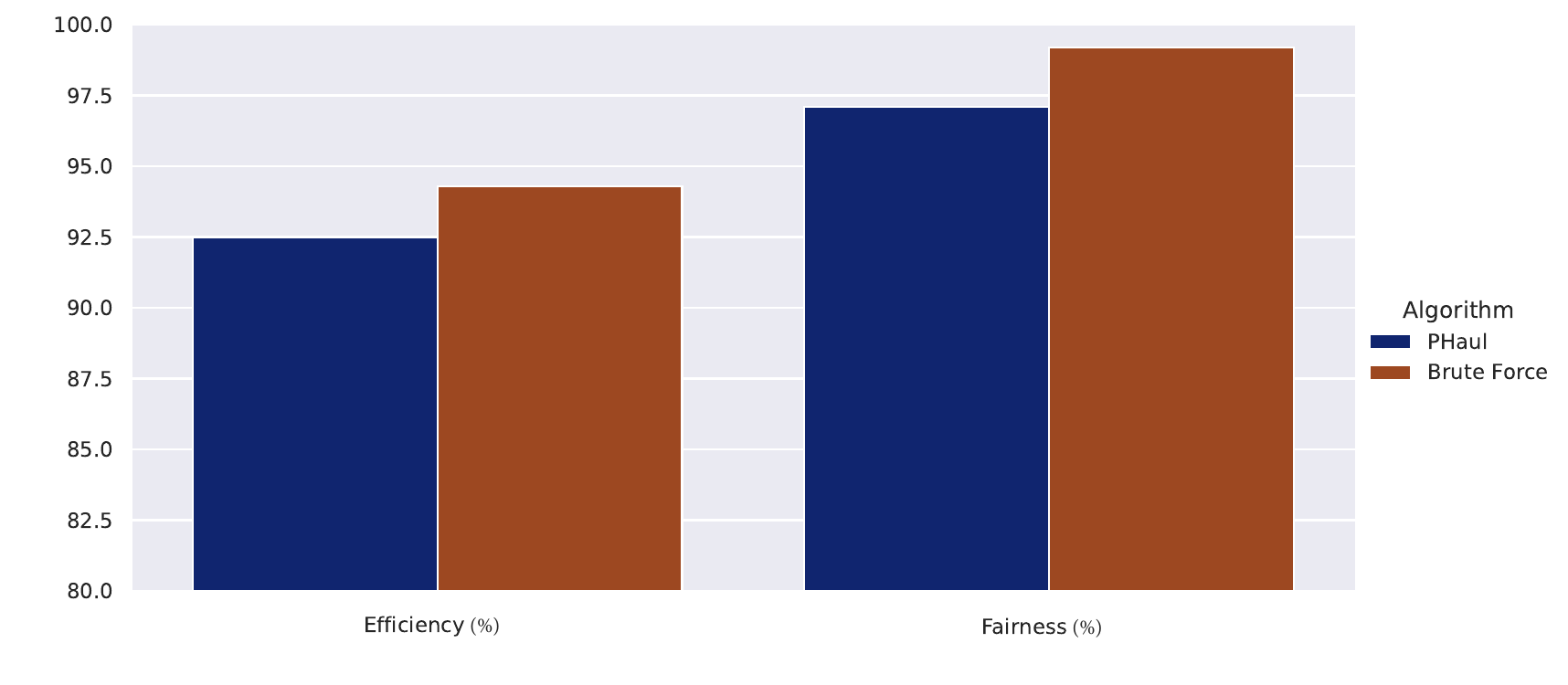}
  \caption{PHaul vs Brute Force for network size 20 and \emph{node\_active\_probability}$=1$}
  \label{fig:phaul_vs_vruteforce}
\end{figure}

To assess how far the PHaul performance lies from the optimal allocation, we compare in Figure \ref{fig:phaul_vs_vruteforce} PHaul against the brute-force agent in terms of efficiency ($\gamma=1$) and fairness ($\gamma=-1$), considering $N^{steps} = 300$ and $training\_steps=20000$. Due to the high computational requirements of the brute force agent we are only able to carry out this benchmark with a limited network size of $20$ IAB nodes and with $K^{max}=1$. To have a meaningful comparison, we need to ensure that the network is saturated, for which we configure \emph{node\_active\_probability}$=1$ and use flow sizes with $\lambda_{min}=500$ Mbps, $\lambda_{max}=750$ Mbps. We can see in Figure \ref{fig:phaul_vs_vruteforce} that both the efficiency and fairness achieved by PHaul lie very close to the brute force agent, which validates the performance of PHaul in the considered scenario. The performance gap between PHaul and brute force is however expected to increase if larger topologies are considered. 

After proving that the PHaul path allocation agent can be effectively trained towards different objective functions, we evaluate next its performance against the subset-sum and radom path allocation agents for a variety of network sizes. 
 
\subsection{Inference mode evaluation against competing heuristics}
\label{subsec:pe_path_alloc_inference}

\begin{figure*}[ht]
    \centering

    \subfigure[Flow size $\lambda_{min}=250$ Mbps, $\lambda_{max}=500$ Mbps]{
        \includegraphics[width=0.48\textwidth]{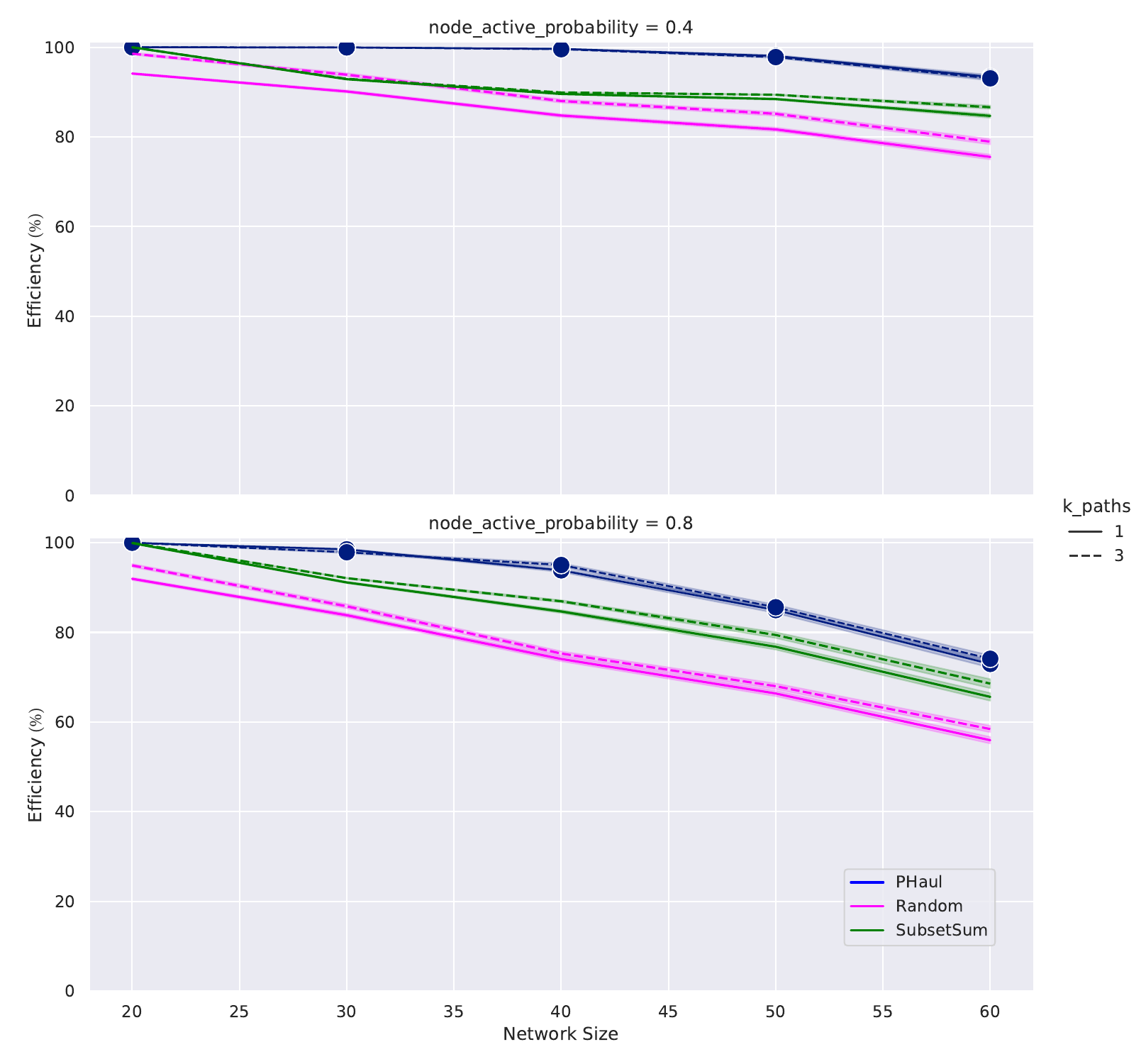}
        \label{fig:subfig1}
    }
    \hfill
    \subfigure[Flow size $\lambda_{min}=500$ Mbps, $\lambda_{max}=750$ Mbps]{
        \includegraphics[width=0.48\textwidth]{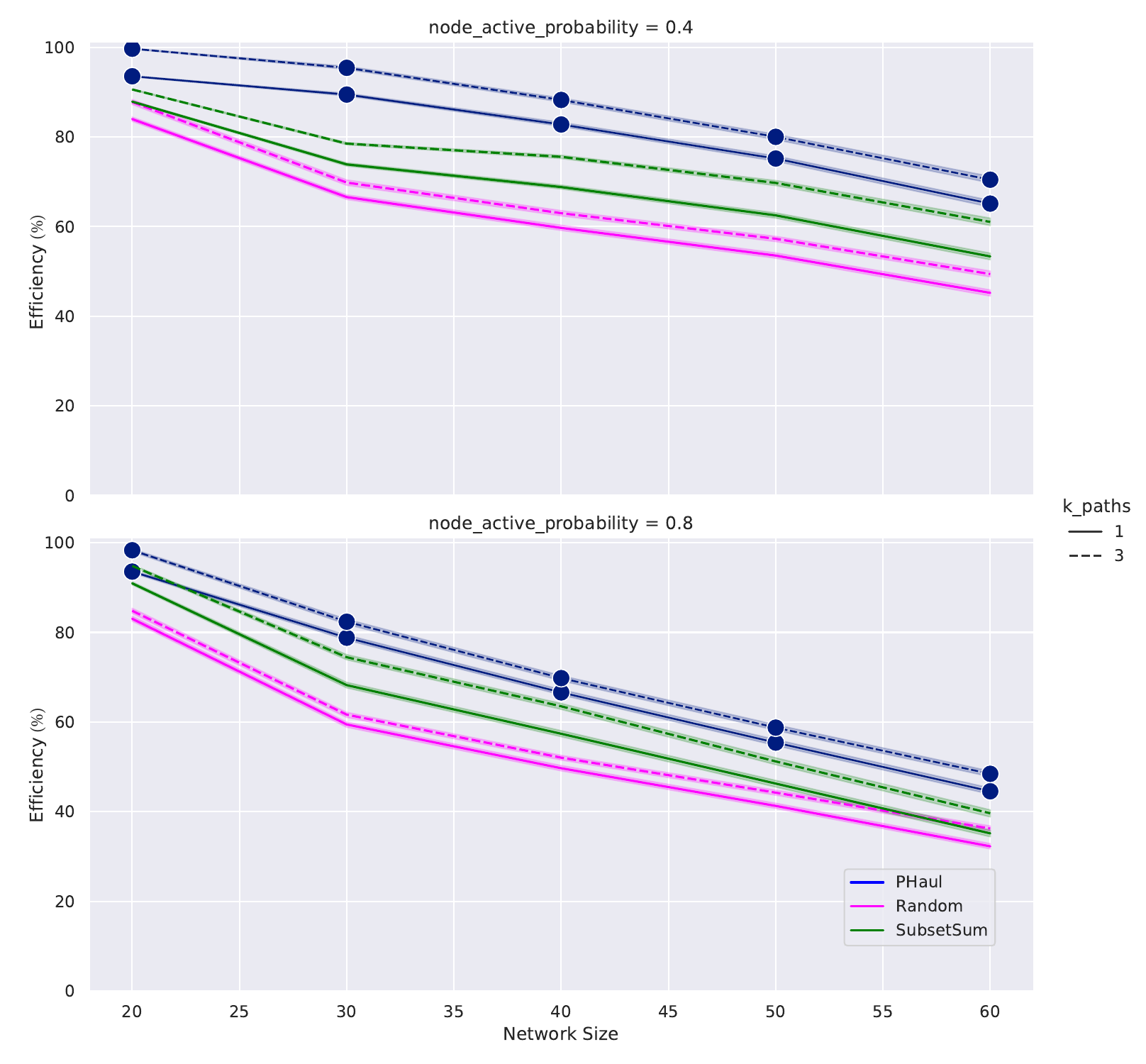}
        \label{fig:subfig2}
    }

    \caption{Efficiency ($\gamma=1$) with a growing network size}
    \label{fig:efficiency_inference}
\end{figure*}

\begin{figure*}[h]
    \centering

    \subfigure[Flow size $\lambda_{min}=250$ Mbps, $\lambda_{max}=500$ Mbps]{
        \includegraphics[width=0.48\textwidth]{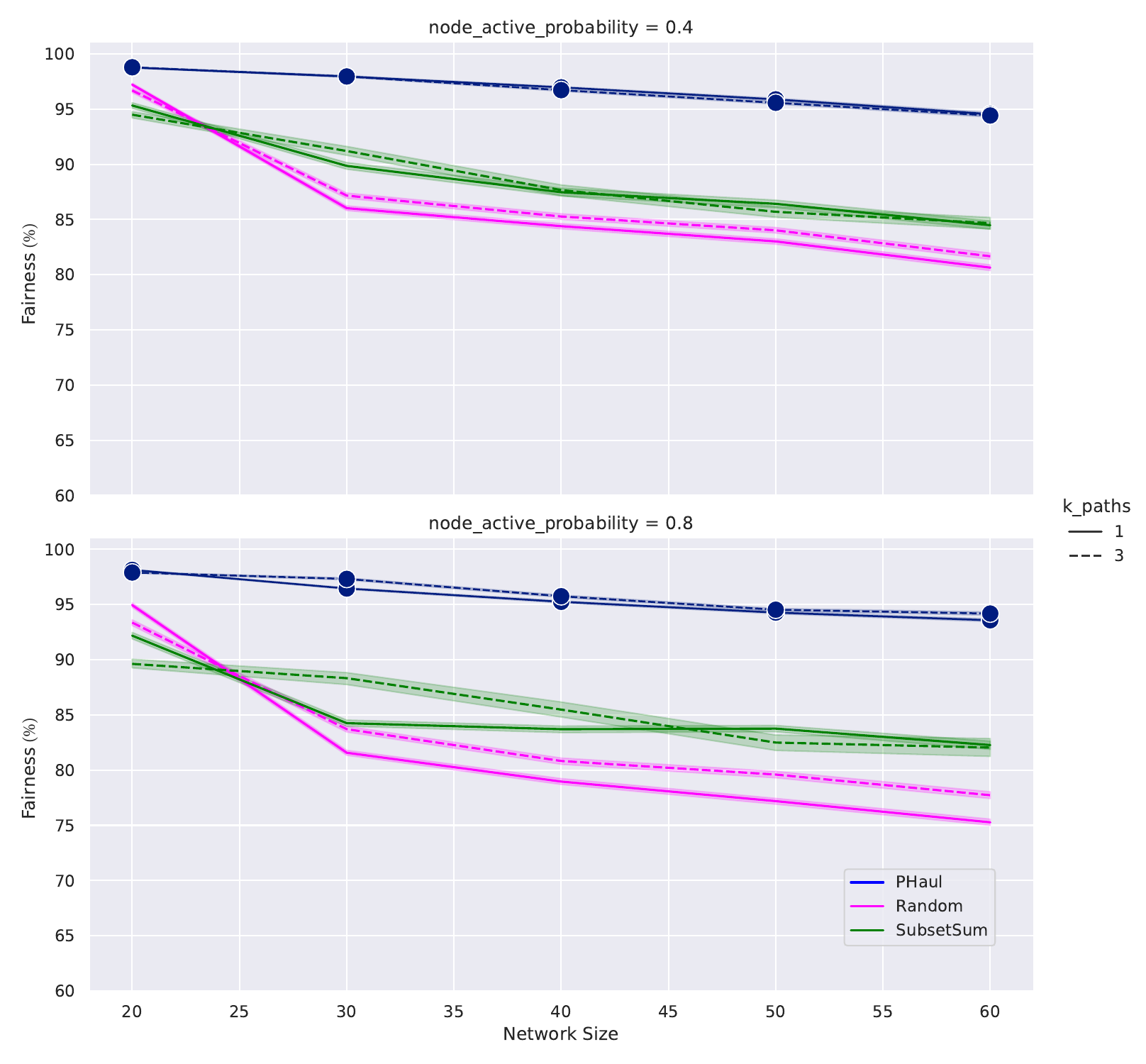}
        \label{fig:subfig1}
    }
    \hfill
    \subfigure[Flow size $\lambda_{min}=500$ Mbps, $\lambda_{max}=750$ Mbps]{
        \includegraphics[width=0.48\textwidth]{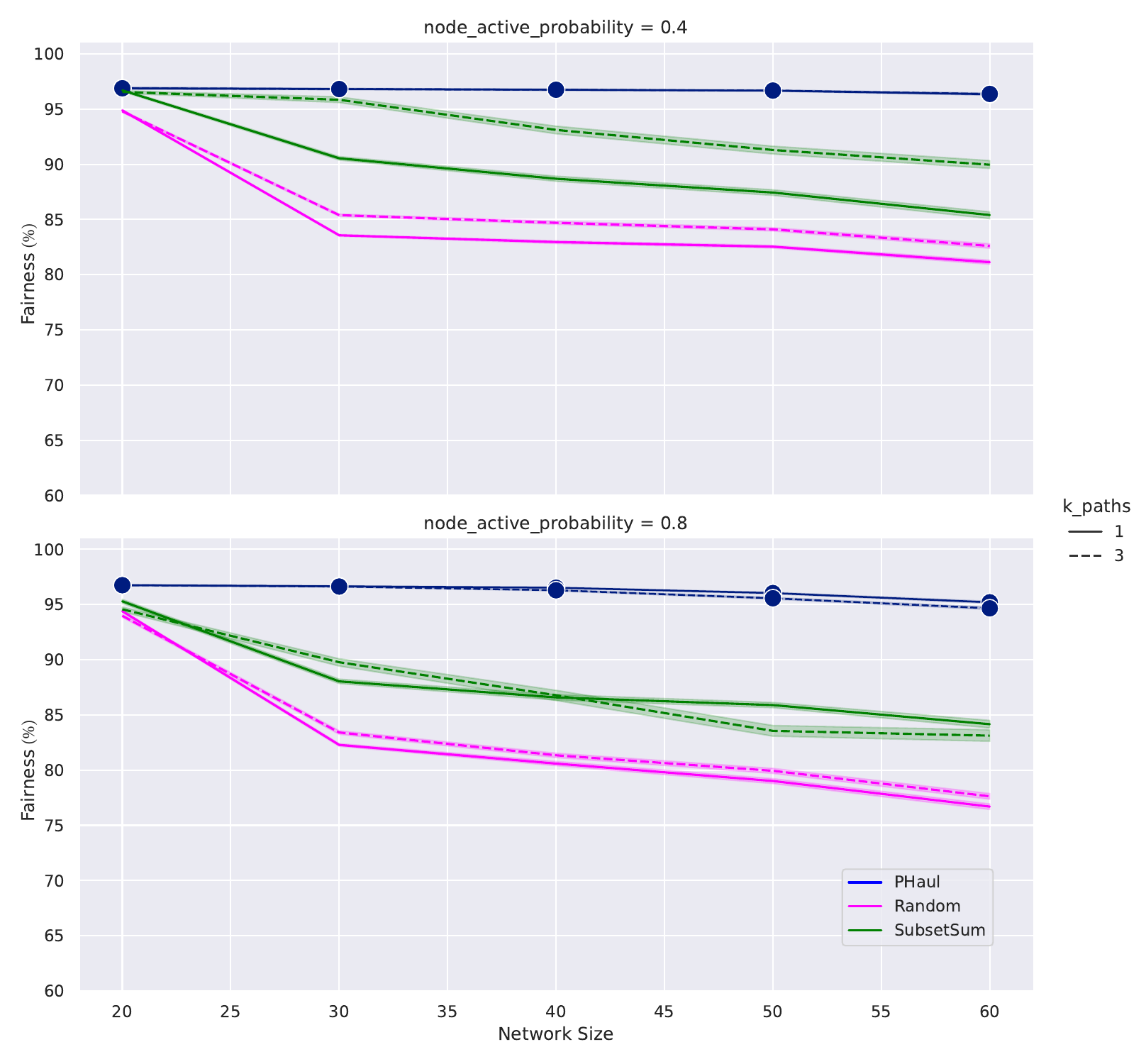}
        \label{fig:subfig2}
    }

    \caption{Fairness ($\gamma=-1$) with a growing network size}
    \label{fig:fairness_inference}
\end{figure*}

Next, we evaluate the performance of the PHaul, the subset-sum and the random path allocation agents, when increasing the size of the IAB network from 20 to 60 nodes, including 3 IAB donor nodes. We do not consider the brute force agent in this section because of its excessive computational time. To consider different load conditions in the network, we set \emph{node\_active\_probability} to $0.4$ and $0.8$ and we evaluate two types of input traffic matrix distributions with a growing flow size in Mbps, namely: i) $\lambda_{min}=250$, $\lambda_{max}=500$, and ii) $\lambda_{min}=500$, $\lambda_{max}=750$. All agents are evaluated for $K^{max}=1$ path in the mm-wave and Sub6 networks, and for $K^{max}=3$ paths in the mm-wave and Sub6 networks.

Figure \ref{fig:efficiency_inference} depicts efficiency ($\gamma=1$) when the network size grows between 20 and 60 IAB nodes, with $node\_active\_probability=0.4$ in the upper row, $node\_active\_probability=0.8$ in the lower row, flow size in Mbps between $\lambda_{min}=250$ and $\lambda_{max}=500$ in the left column and flow size in Mbps between $\lambda_{min}=500$ and $\lambda_{max}=750$ in the right column.

We can see how for all network configurations the PHaul agent outperforms subset-sum, which in turn outperforms the random agent. A maximum gain of 17\% is observed for PHaul when compared to subset-sum, and of 36\% when compared to random. All agents benefit from considering a larger number of paths, but this gain is more evident when the flow data rates are higher ($\lambda_{min}=500$, $\lambda_{max}=750$). Note that subset-sum is a well proven bin-packing heuristic that sorts backhaul flows in decreasing order of size and greedily starts allocating them one at a time. The reason why PHaul is able to outperform subset-sum, is that in the training process PHaul is able to learn a representation of the topology of the IAB network, which it can then correlate with a given traffic matrix distribution to derive non-trivial allocations that result in good performance. For all the agents efficiency decreases as network size increases, and the decrease is higher for higher flow data rates. The reason for this behavior is that introducing new IAB nodes results in a higher offered load, regulated by the parameter \emph{node\_active\_probability}. This effect dominates over any increase in cross-section bandwidth that may result from the additional backhaul links contributed by the new IAB nodes.

Figure \ref{fig:fairness_inference} depicts now the results in terms of fairness ($\gamma=-1$), for the same experiments described in Figure \ref{fig:efficiency_inference}. Unlike efficiency, fairness exhibits a rather flat behavior when the size of the IAB network grows. When the IAB network grows, the number of backhaul flows increases and the effective bandwidth allocated to each flow decreases. In our network model, each IAB node allocates per-flow capacities in a bottleneck link using a water-filling algorithm. Therefore, the means that the different agents have to improve fairness is to select the paths for each flow such that all flows in the network achieve a similar effective capacity. We can see in Figure \ref{fig:fairness_inference} how PHaul is the best agent in achieving a fair allocation, resulting in a close to perfect fairness for all considered network settings, and for both 1 and 3 available paths. A maximum gain of 13\% is observed in terms of fairness for PHaul when compared to subset-sum, and of 20\% when compared to random. The reason for the good performance of PHaul, is that in the training phase PHaul is able to learn correlations between a given traffic matrix and the set of available paths that result in a good fairness metric.

After verifying in this section that PHaul consistently offers a superior performance in terms of throughput efficiency and fairness than subset-sum and random, we study next the performance of the different agents in terms of execution time.

\subsection{Benchmarking PHaul Execution Time}
\label{subsec:pe_path_alloc_exec_time}

The main goal of PHaul, as stated in Section \ref{sec:agent_design}, is to periodically read the traffic matrix from the physical network to then update the mapping between backhaul flows and pre-provisioned backhaul paths. The frequency of these updates is thus limited by the execution time of the path allocation agent.

\begin{figure*}[ht]
    \centering

    \subfigure[Efficiency ($\gamma=1$) for $K^{max}=1$ and $K^{max}=3$]{
        \includegraphics[width=0.47\textwidth]{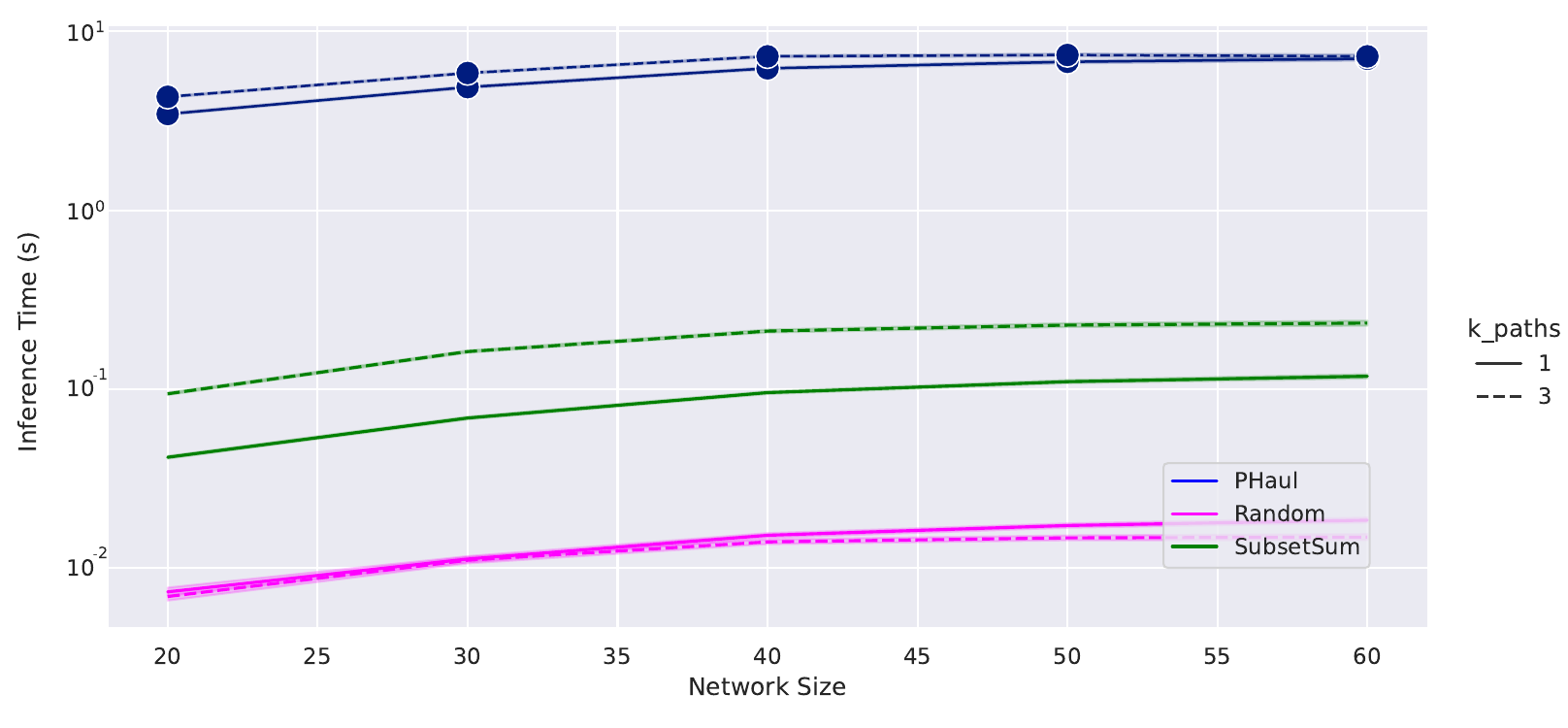}
        \label{fig:subfig1}
    }
    \hfill
    \subfigure[Fairness ($\gamma=-1$) for $K^{max}=1$ and $K^{max}=3$]{
        \includegraphics[width=0.47\textwidth]{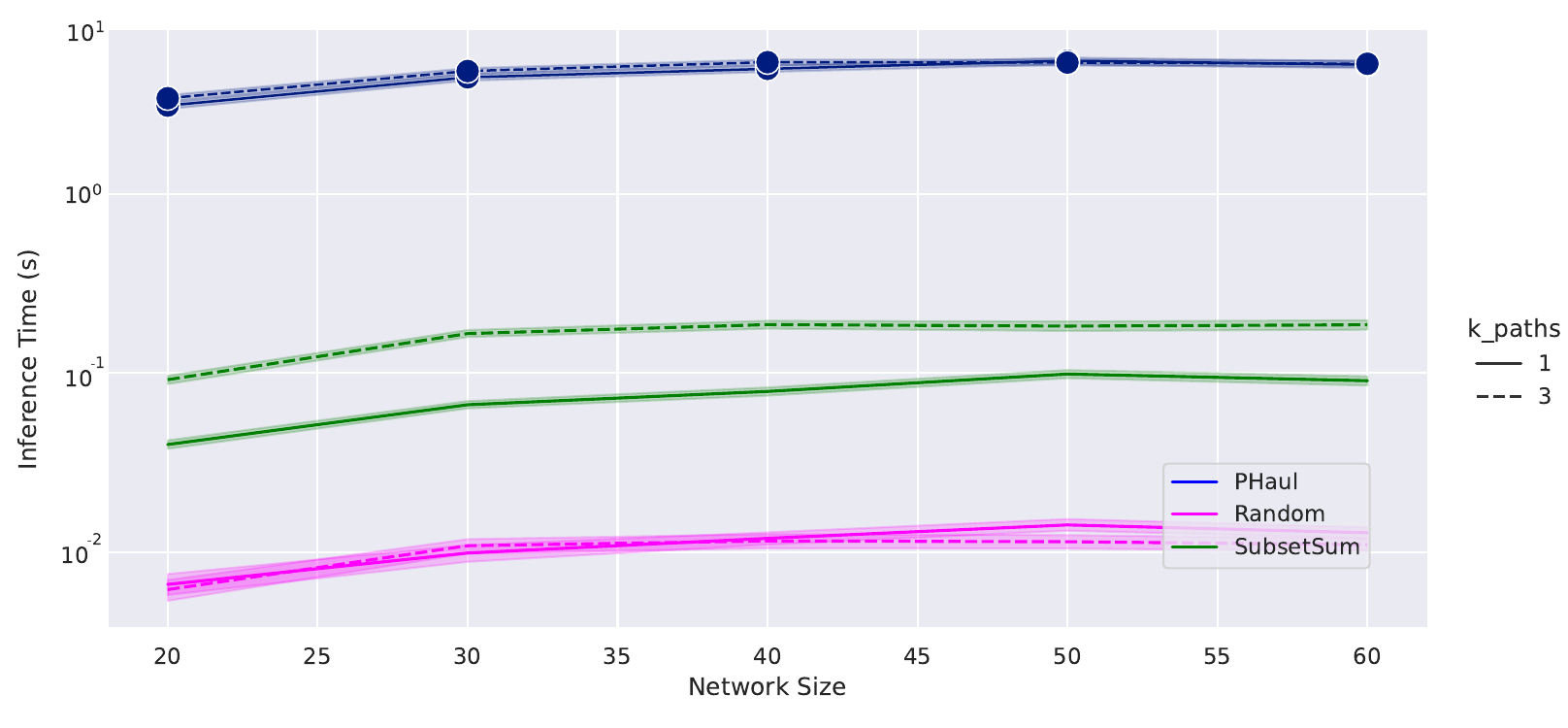}
        \label{fig:subfig2}
    }

    \caption{Inference time (secs)}
    \label{fig:pe_exec_time}
\end{figure*}

Figure \ref{fig:pe_exec_time} depicts the average execution time of the PHaul, subset-sum and random agents, when increasing the IAB network between 20 and 60 nodes. The left hand size depicts the execution times when optimizing for efficiency, which result from averaging all the network configuration considered in Figure \ref{fig:efficiency_inference}. 
The right hand side depicts the corresponding execution time when optimizing for fairness, considering the network configurations included in Figure \ref{fig:fairness_inference}.

The execution times reported Figure \ref{fig:pe_exec_time} are of course dependent on the platform where the agents are run, which in our case consist of a single Intel Xeon E5-2618L v4 CPU. Notice that in a real implementation the PHaul path allocation agent can be implemented in a centralized location where enough compute resources are available. 
We can see in Figure \ref{fig:pe_exec_time} how, as expected, the execution times of PHaul are larger than those of subset-sum and random. However, PHaul keeps an execution time below 10 seconds that slightly increases with the network size. A 10 second interval to reconfigure the forwarding in the backhaul is a reasonable value  because UEs connected to a cell transition from RRC\_CONNECTED to RRC\_IDLE after an inactivity timeout of around 10 seconds \cite{sodalite_tnsm}, which means that sampling at this frequency is adequate to observe significant changes in the traffic matrix.

It is also relevant to observe how the execution time of PHaul is fairly independent of the network size. The reason is that the execution time of PHaul is dominated by the $N^{steps}$ parameter, which defines the number of interactions with the network digital twin (c.f. Section \ref{sec:agent_design}). It is thus possible to reduce execution time in PHaul by reducing $N^{steps}$, at the cost of losing accuracy in terms of the objective function, as depicted in Figure \ref{fig:phaul_training_eval}. Regarding the number of paths $K^{max}$, we can see that they only have a marginal impact on the execution time of PHaul. The reason is that increasing $K^{max}$ translates into an increase in the size of the action space, which may impact convergence time in the training phase, but it results in a minor impact with respect to the time required to decide what action to choose in inference phase. This is not the same for subset-sum, which is clearly affected by the number of paths, as it needs to evaluate $2K^{max}$ candidate path allocations for each flow. Finally, the execution time of the random agent is negligible, as it only involves computing a random number.

\begin{figure*}[ht]
    \centering

    \subfigure[Efficiency ($\gamma=1$) for $K^{max}=1$ and $K^{max}=3$]{
        \includegraphics[width=0.48\textwidth]{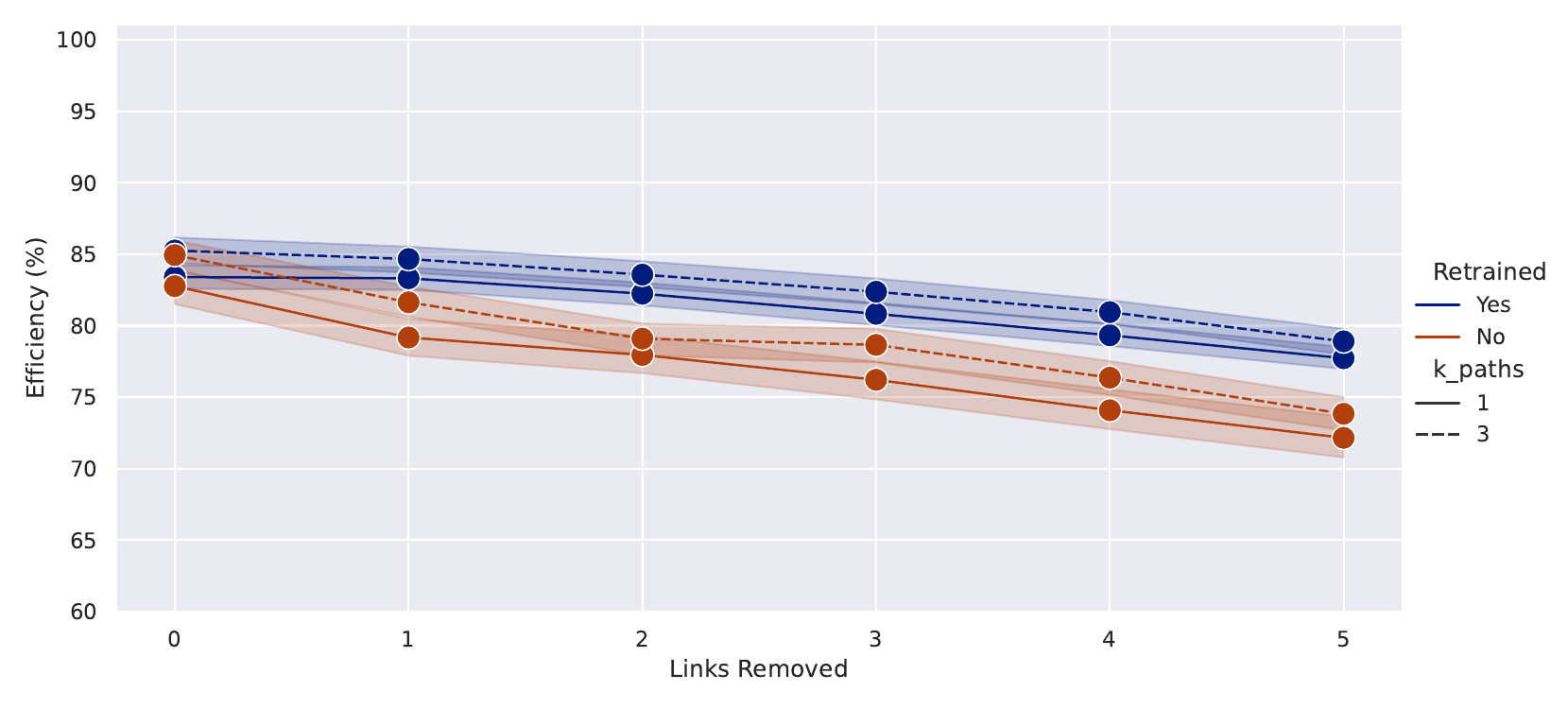}
        \label{fig:pe_link_eff}
    }
    \hfill
    \subfigure[Fairness ($\gamma=-1$) for $K^{max}=1$ and $K^{max}=3$]{
        \includegraphics[width=0.48\textwidth]{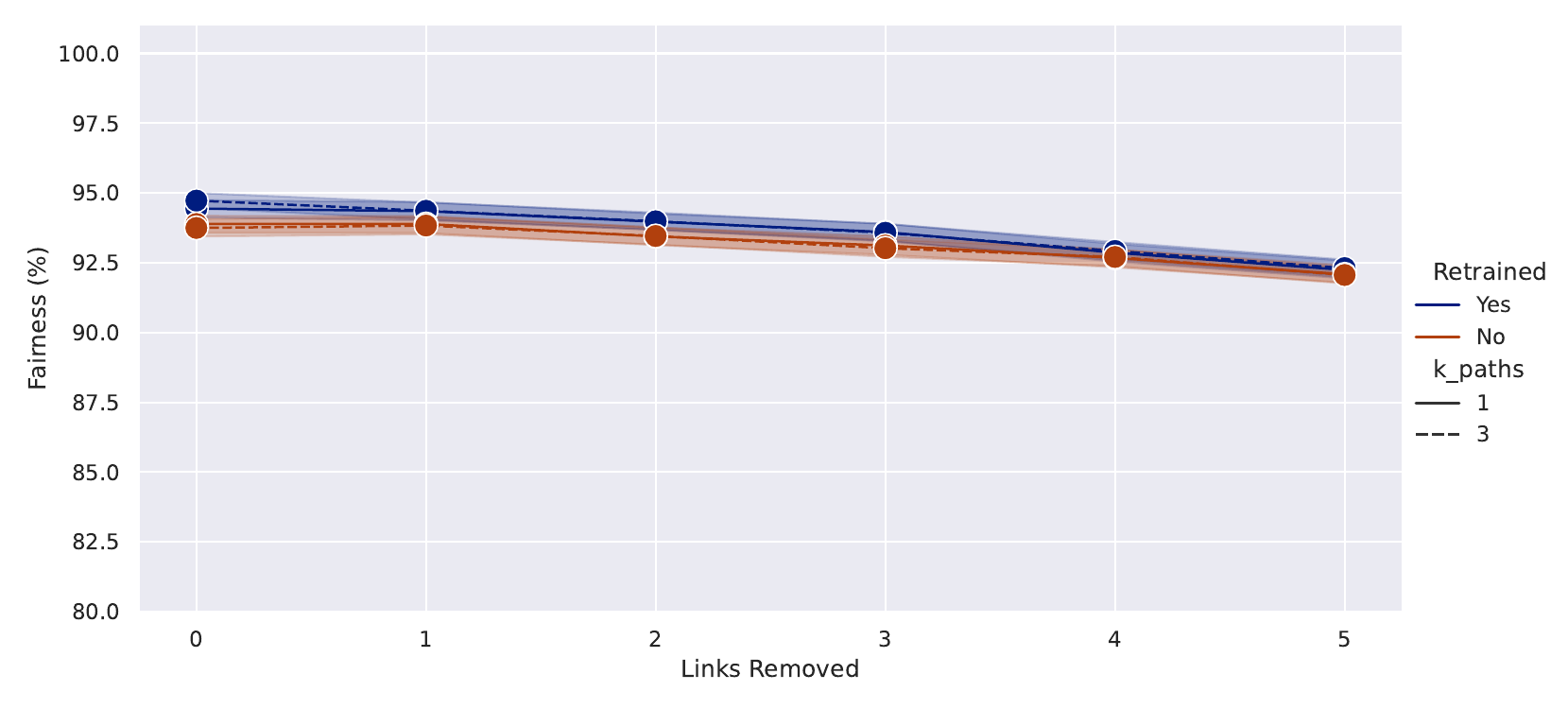}
        \label{fig:pe_link_fair}
    }

    \caption{\textcolor{black}{Impact of retraining when removing links from the network, for flow size ($\lambda_{min}=500$ Mbps, $\lambda_{max}=750$ Mbps) and $node\_active\_probability=0.4$}}
    \label{fig:pe_link_breaks}
\end{figure*}

\subsection{\textcolor{black}{Impact of broken links}}\label{subsec:pe_path_alloc_broken}
\textcolor{black}{Given the nature of the IAB wireless backhaul, transitory broken links can occur due to link blockage at mm-wave frequencies, or due to unplanned interference at Sub6 frequencies. The goal of this section is to evaluate how resilient is PHaul to these events, since it is not realistic to assume that PHaul can be retrained every time a link breaks. Upon a broken link, the internal IAB routing protocol will restore end-to-end connectivity by re-routing backhaul flows. Notice that the PHaul action space simply assigns for each flow a path defined by a path index. Thus, if upon a link break the actual hop-by-hop sequence of a path is modified, but the path index is maintained and the path still connects the same source and destination, then PHaul is able to continue forwarding packets through that path. Nevertheless, the resulting topology upon a link break will differ from the topology that PHaul has been trained on, and hence a performance degradation can be expected. The goal of this section is to quantify this degradation.}

\textcolor{black}{Figure \ref{fig:pe_link_breaks} depicts the results of an experiment where we consider a network of 40 IAB nodes, with flow size $\lambda_{min}=500$ Mbps, $\lambda_{max}=750$ Mbps and $node\_active\_probability=0.4$, and increase the number of simultaneous broken links from 1 to 5, which we consider representative of this network size. For each point in the x-axis we consider 10 different topologies and 250 random samples with different traffic matrixes. In each sample we randomly remove $x$ links from the network, but ensure that end-to-end connectivity for all backhaul flows remains possible. Figure \ref{fig:pe_link_breaks} compares the performance in terms of efficiency ($\gamma=1$) and fairness ($\gamma=-1$), considering the ideal performance where PHaul is retrained every time that a link is removed from the topology (shown in blue), versus the performance to be expected in practice when PHaul has only been trained for the full topology but continues to perform inferences when links are removed (shown in brown).}

\textcolor{black}{Looking at Figure \ref{fig:pe_link_eff} we see that overall efficiency reduces as the number of broken links increases, because the network has less capacity, but the efficiency loss because of having PHaul operate over a network with broken links is only around 5\%, both for $K^{max}=1$ and $K^{max}=3$ paths. Looking at Figure \ref{fig:pe_link_fair} we observe that the loss in fairness is even smaller, being around 2\%. We note that the tree-like structure of IAB backhaul networks tends to result in backup paths that are similar to the original ones, which helps explain the good performance of PHaul observed in this experiment.}

\textcolor{black}{In the next section we continue analyzing the resilience of PHaul to untrained topologies that significantly differ from the topology PHaul has been trained on.}

\subsection{\textcolor{black}{Impact of untrained topologies and Sub6 spectrum}}
\label{subsec:pe_path_alloc_untrained_topos}

\begin{figure*}[ht]
    \centering

    \subfigure[Efficiency ($\gamma=1$) for $K^{max}=1$ and $K^{max}=3$]{
        \includegraphics[width=0.48\textwidth]{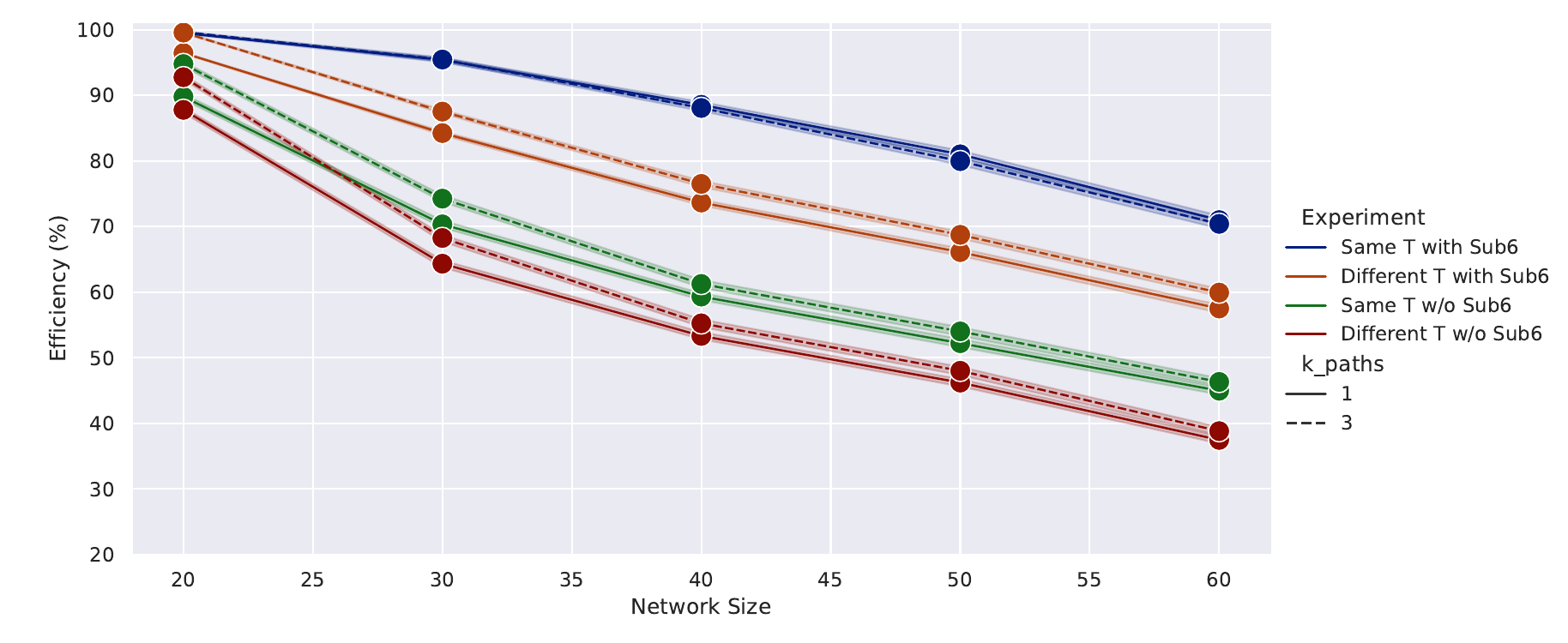}
        \label{fig:pe_untrained_sub6_eff}
    }
    \hfill
    \subfigure[Fairness ($\gamma=-1$) for $K^{max}=1$ and $K^{max}=3$]{
        \includegraphics[width=0.48\textwidth]{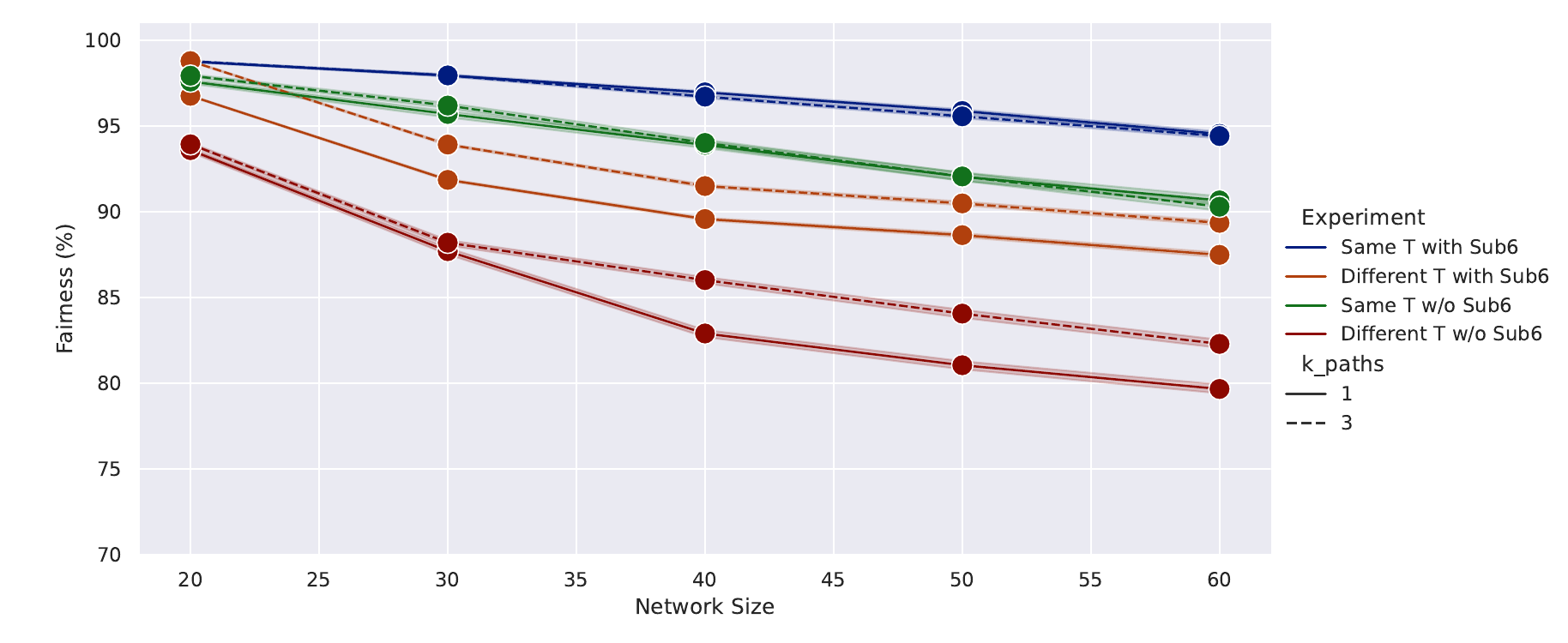}
        \label{fig:pe_untrained_sub6_fair}
    }

    \caption{Impact of untrained topologies and Sub6 spectrum, for flow size ($\lambda_{min}=500$ Mbps, $\lambda_{max}=750$ Mbps) and $node\_active\_probability=0.4$}
    \label{fig:pe_untrained_topologies}
\end{figure*}

\textcolor{black}{Having seen in the previous section that PHaul is reliable to small changes in the topology, in this section we deepen our evaluation in two directions. First, we evaluate the performance of PHaul on topologies that differ significantly from the topology that PHaul has been trained on. Second, we evaluate the performance of PHaul with and without Sub6 connectivity, to quantify the gains that adding Sub6 spectrum provides on efficiency and fairness metrics.}

\textcolor{black}{To evaluate PHaul with untrained topologies, we consider again 10 topologies for each considered network size, but train PHaul with only one of the 10 topologies. All topologies share the same number of IAB nodes and the same number of donor nodes, which is what makes PHaul applicable across them. Topologies though, differ in the number of layers and in the links connecting IAB nodes. To evaluate the gains provided by Sub6 spectrum, we run each topology under two configurations: i) a first configuration where both mm-wave and Sub6 spectrum are available, and ii) a second configuration where only mm-wave links are available.}

\textcolor{black}{Figure \ref{fig:pe_untrained_topologies} depicts for a growing network size the performance in terms of efficiency with $\gamma=1$ (left graph) and fairness with $\gamma=-1$ (right graph). For this experiment we assume flow size $\lambda_{min}=500$ Mbps, $\lambda_{max}=750$ Mbps and $node\_active\_probability=0.4$, and consider $K^{max}=1$ and $3$. We consider the following four configurations: i) Sub6 enhanced IAB where PHaul is trained specifically for each topology (blue lines), ii) Sub6 enhanced IAB where PHaul is only trained for one topology (brown lines), iii) mm-wave only IAB where PHaul is trained for each topology (green lines), and iv) mm-wave only IAB where PHaul is only trained for one topology (red lines). For each configuration we consider $K^{max}=1$ and $K^{max}=3$ paths.}

\textcolor{black}{Looking at the impact in efficiency in Figure \ref{fig:pe_untrained_sub6_eff}, we can see how using untrained topologies results in a slight degradation of about 9\% in efficiency. Removing Sub6 spectrum, but retraining PHaul every time, results in an efficiency loss of around $25\%$, which justifies the gains provided by Sub6 spectrum, since, using PHaul, a Sub6 enhanced IAB network with untrained topologies is still more efficient that a perfectly trained mm-wave only IAB network. Finally, removing Sub6 spectrum and using untrained topologies results in an overall degradation slightly above $30\%$.}

\textcolor{black}{Looking at fairness in Figure \ref{fig:pe_untrained_sub6_fair} we can see that the trend is maintained, but interestingly, removing Sub6 spectrum results in a degradation of around $4\%$, whereas using untrained topologies result in a higher degradation of around $7\%$. The reason is that when removing Sub6 spectrum the overall network capacity reduces, but PHaul is still able to allocate the available capacity across flows in a fair way. Fairness is however impacted when PHaul runs over untrained topologies, although the impact is small. Finally, removing Sub6 spectrum and using untrained topologies results in the worst case ($K^{max}=1$) in a degradation of around $15\%$.}

\textcolor{black}{These results quantify the benefits of adding Sub6 spectrum to the IAB backahul, as proposed in this paper, and validate that PHaul has a graceful degradation when used over topologies that differ significantly from the topologies that have been used for training, which validate the application of PHaul in practical networks.}

\section{Conclusions}
\label{sec:conclusions}

Future bandwidth-hungry 6G services will require the deployment of wireless access networks providing gigabit capacities. A key challenge towards the design of 6G is therefore the development of network architectures that can provide the required capacity while being economically sustainable. 3GPP IAB is a novel network architecture that addresses this concern by allowing to reuse the same spectrum in access and backhaul, thus minimizing the need of expensive fibre layouts when deploying a new radio access network.

In this paper we have proposed to enhance 3GPP IAB networks by adding Sub6 spectrum in the backhaul segment, where the greenfield 6 GHz band, validated in WRC-23 for IMT use \cite{wrc23}, could be used for this purpose. In addition to the added backhaul capacity, the Sub6 spectrum complements mm-wave in terms of coverage, which is critical in dense urban environments. Based on our Sub6 enhanced IAB architecture, we have proposed PHaul, which is a novel DRL-based forwarding agent for IAB networks that adapts the paths used to forward the backhaul flows according to a periodically sampled traffic matrix.

Using a simulation approach that models realistic IAB topologies, we have analyzed the training properties of PHaul and have evaluated its performance in terms of throughput efficiency and fairness with respect to two competing IAB forwarding agents. In our experiments PHaul always outperforms competing approaches, demonstrating gains of up to 36\% in terms of efficiency and of up to 20\% in terms of fairness. We have shown that the PHaul is able to execute path allocation inferences in less than 10 seconds, which is a reasonable frequency to be applied to real IAB networks. Finally, we have shown that PHaul degrades smoothly when there are differences between the IAB topology and the topology that PHaul has been trained on, and we have explicitly quantified the performance gain that is due to adding Sub6 capabilities to the IAB network. 

As future work we consider the comparison of different DRL algorithms applied to the PHaul forwarding agent, as well as the evaluation of additional traffic engineering criteria, for example incorporating criteria related to energy efficiency.


\begin{IEEEbiography}[{\includegraphics[width=1in,height=1.25in,clip,keepaspectratio]{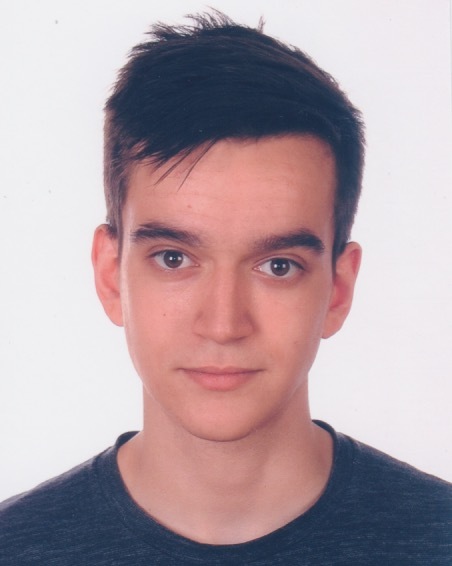}}]{Jorge Pueyo} is a research and development engineer at i2CAT (Barcelona, Spain). He holds a Bachelor's degree in Telecommunication Technologies and Services Engineering from the Polytechnic University of Catalonia (UPC) and a Master's degree in Advanced Telecommunications Technologies, also from UPC. He has extensive experience in software development, primarily using Python and Java, as well as expertise in cloud development, data analysis, and machine learning. Currently, he is actively involved in various European projects related to O-RAN technologies.
\end{IEEEbiography}
\vspace{-1.5cm}

\begin{IEEEbiography}[{\includegraphics[width=1in,height=1.25in,clip,keepaspectratio]{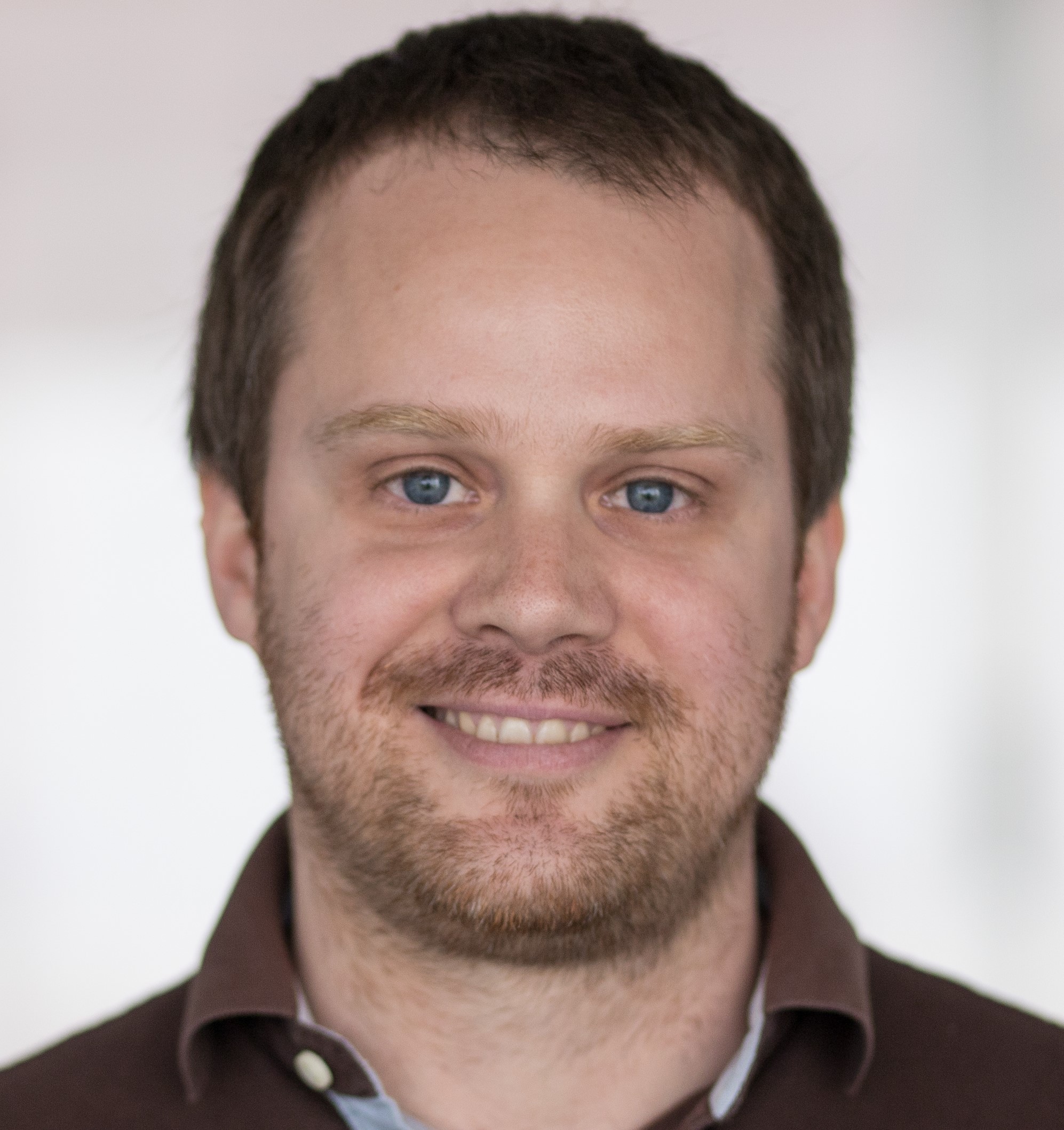}}]{Daniel Camps-Mur} is currently leading the Mobile and Wireless Internet (MWI) group at I2CAT in Barcelona, Spain. Previously, Daniel was a senior researcher at NEC Network Laboratories in Heidelberg, Germany. In 2004 he received a Master’s degree and in 2012 a Ph.D. degree from the Polytechnic University of Catalonia (UPC). His research interests include mobile networks, software defined networking and communications protocols for the Internet of Things.
\end{IEEEbiography}
\vspace{-1.2cm}

\begin{IEEEbiography}[{\includegraphics[width=1in,height=1.25in,clip,keepaspectratio]{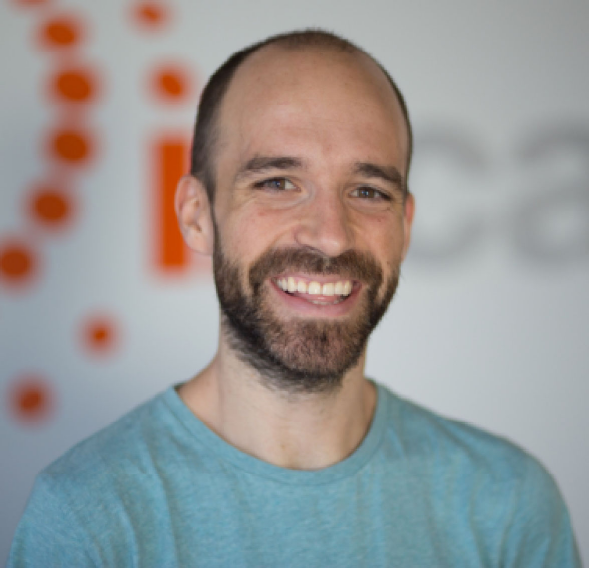}}]{Miguel Catalan-Cid} is a senior research engineer at i2CAT in Barcelona, Spain. He holds since 2008 a Master’s degree and since 2016 a Ph.D. from the Polytechnic University of Catalonia (UPC). He has an extensive experience in wireless networks, analysis and definition of communication protocols, utilisation of simulation tools, and programming embedded systems and micro-controllers. He is currently being involved in different H2020 projects related to B5G and O-RAN technologies.  
\end{IEEEbiography}
\vspace{-1.2cm}

\end{document}